\documentclass[intlimits,twoside,a4paper]{article}
\usepackage{wrapfig}
\usepackage{gensymb}
\usepackage{graphicx}

\usepackage[cp1251]{inputenc}

\usepackage[eqsecnum]{cmpj3}

\newcommand{\be}{\begin{equation}}
\newcommand{\ee}{\end{equation}}
\newcommand{\bea}{\begin{eqnarray}}
\newcommand{\eea}{\end{eqnarray}}

\issue{2017}{20}{2}{23706}
\doinumber{10.5488/CMP.20.23706}

\title[Influence of electric fields]{Influence of electric fields on dielectric properties of GPI ferroelectric}
\author[I.R. Zachek, R.R. Levitskii, A.S. Vdovych, I.V. Stasyuk]{I.R. Zachek\refaddr{label1}, R.R. Levitskii\refaddr{label2}, A.S. Vdovych\refaddr{label2}, I.V. Stasyuk\refaddr{label2}
}
\addresses{\addr{label1} Lviv Polytechnic National
University, 12 Bandera St., 79013 Lviv, Ukraine
\addr{label2} Institute for Condensed Matter Physics of the
National Academy of Sciences of Ukraine,\\ 1 Svientsitskii St.,
79011 Lviv, Ukraine }

\date{Received April 20, 2017, in final form May 23, 2017}
\DeclareMathOperator{\Sp}{Sp}

\begin{document}

\maketitle

\begin{abstract}

Using modified microscopic model of GPI by taking into account the piezoelectric coupling with strains
 $\varepsilon_i$ in the frames of two-particle cluster approximation,
the components of polarization vector and static dielectric permittivity tensor of the crystal at applying the external transverse electric fields  $E_1$ and $E_3$ are calculated. An analysis of the influence of these fields on thermodynamic characteristics of GPI is carried out.
 A satisfactory quantitative description of the available experimental data for these characteristics has been obtained at a proper choice of the model parameters.

\keywords ferroelectrics, electric field, polarization, dielectric permittivity, phase transition
\pacs 77.22.-d, 77.22.Ch, 77.22.Ej, 77.65.-j, 77.80.Bh
\end{abstract}

\section{Introduction}

One of the actual problems in physics of ferroelectric materials is the study of the effects that appear under the action of an external electric field. It can be a powerful tool for purposeful control of their physical characteristics. The effects of the action of external fields depend both on the intensity and  the type of such an action, and on the properties of the materials. The application of an electric field is a very important instrument for the investigation of ferroelectric materials with a complex spatial arrangement of the local effective dipole moments.
Consequently, phase transitions with different order parameters connected with each other can take place in these materials. In particular, it appears  possible to influence this system by means of an electric field, which is perpendicular to a spontaneous polarization, and to study the changes of polarization and the other dielectric properties.

\begin{figure}[!t]
\begin{center}
\includegraphics[scale=0.7]{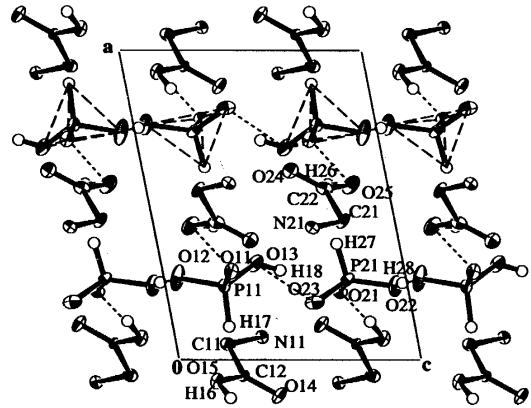}
\end{center}
\vspace{-3mm}
\caption{The lattice structure of glycinium phosphite crystal \cite{Averbuch1993}.} \label{gpi_str2}
\end{figure}

One of the most interesting examples of a crystal sensitive to an electric field effect is the glycinium phosphite (GPI), which belongs to ferroelectric materials with hydrogen bonds \cite{dac,Baran1996}. At the room temperature this crystal has a monoclinic structure (space group P$2_{1}$/a) \cite{Averbuch1993}. The hydrogen bonds between the tetrahedra HPO$_{3}$ form infinite chains along the crystallographic
$c$-axis (figure~\ref{gpi_str2}).
There are two types of hydrogen bonds with the length $\sim 2.48$~{\AA} and $\sim 2.52$~{\AA}
\cite{Averbuch1993,Shikanai2002,Taniguchi2003}. The ordering of protons on these bonds
\cite{Shikanai2002,Taniguchi2003} causes an antiparallel orientation of the components of dipole moments of the equivalent hydrogen bonds along the crystallographic axes
$a$ and $c$ in the neighbouring chains. However, the changes of the distances between ions in the tetrahedra HPO$_{3}$ and the parallel ordering of the corresponding components of dipole moments along the $b$-axis in the chains causes a total dipole moment along this axis.

Consequently, at the temperature 225~K the crystal passes to the ferroelectric state  (space group P2{$_1$})
with a spontaneous polarization perpendicular to the chains of hydrogen bonds. It is necessary to note that the phase transition in GPI is closely connected with the short- and long-range interactions within these chains and between them.

The study of the effect of deuteration on $T_{\text c}$ witnesses  in favour of the proton ordering mechanism of a phase transition due to a strong isotopic shift of the transition temperature ($T_{\text c}^{\text D}-T_{\text c}^{\text H} = 97$~K \cite{Baran1997}).

The results of measuring the frequency dependence of dielectric permittivity
\cite{Tchukvinskyi1997,Sobiestianskas1998} also testify that the  phase transition in this crystal is of the order-disorder type. It should be also mentioned that the data obtained based on the slow neutron scattering investigation, indicate the  reorientations and deformations of the ionic groups (phosphite ions). The revealed temperature anomalies of elastic constants near  $T_{\text c}$ \cite{Furtak1997} manifest an important role of deformation processes in a phase transition in GPI.

Highly important are the investigations of transverse electric fields effects on the physical characteristics of GPI.
A crystal seems to be quite special in this respect. The experiment, carried out in \cite{Stasyuk2003,Stasyuk2004}, showed a unique sensitivity to a transverse field $E_z$. As it was established, such a field, applied to the crystal in ferroelectric phase (at $T<T_{\text c}^0$), is capable of reorienting the local dipole moments that are connected with protons on hydrogen bonds and with adjacent ionic glycine groups.
Consequently, at some critical field  $E_z^{\text c}$ there occurs a phase transition, at which a spontaneous polarization along $OY$-axis disappears and only the component $P_z$ remains. Such an effect resembles the well known spin-flop transition in antiferromagnetics under the action of an external magnetic field. On the other hand, as was shown in \cite{Stasyuk2003,Stasyuk2004}, under the action of the field $E_z$ there occurs a decrease of critical temperature of ferroelectric phase transition proportionally to $E_z^2$. The existence of considerable (and increasing with the field) anomalies of transverse dielectric permittivity $\varepsilon_{zz}$ in the region of transition at $E_z\neq0$ was revealed.

An explanation of the discovered effects was given in \cite{Stasyuk2004} and \cite{Stasyuk2003,Stasyuk2004Ferro} based on the phenomenological Landau theory and within the microscopic model approach, respectively. However, it failed to achieve a full quantitative description of the observed temperature and field behaviour of $\varepsilon_{zz}$, inasmuch as the reasons of a smeared character of such dependences remain unclear.

In the present work we continue the study of the transverse field effect, based on a microscopic description within the model of a deformed crystal \cite{Stasyuk1983}. We supplement the approach, applied in \cite{Stasyuk2003,Stasyuk2004Ferro}, by taking into account the lattice strains and piezoelectric coupling. At the same time, our goal is to consider the wider range of phenomena connected with the action of transverse fields $E_z$ and $E_x$ on a ferroelectric phase transition and on dielectric and piezoelectric characteristics of GPI crystal.

\section{The model}

\begin{figure}[!t]
\begin{center}
\includegraphics[scale=0.6]{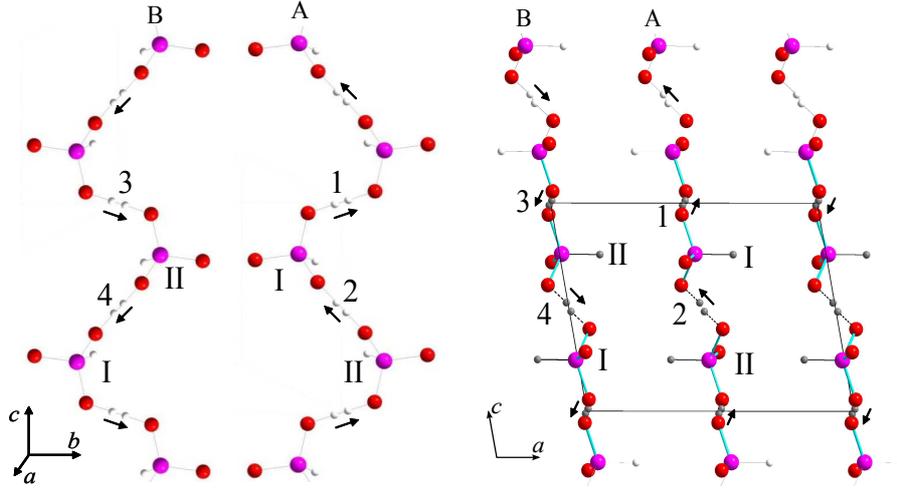}
\end{center}
\caption{(Color online) Orientations of vectors ${\vec d}_{qf}$ in the primitive cell
in the ferroelectric phase.} \label{struktura}
\end{figure}

We consider a system of protons in GPI, localised on O--H$\ldots${O} bonds, which form zigzag chains along the $c$-axis of a crystal.
Dipole moments ${\vec{d}}_{qf}$ ($q$ is a number of a primitive cell, $f=1,\dots,4$) are ascribed to the protons on the bonds. In the ferroelectric phase, the dipole moments compensate each other  (${\vec d}_{q1}$ with ${\vec d}_{q3}$, ${\vec d}_{q2}$ with ${\vec d}_{q4}$) in directions  $Z$ and $X$, and simultaneously supplement each other in the direction $Y$, creating a spontaneous polarization. Vectors ${\vec d}_{qf}$ are oriented at some angles to crystallographic axes and have longitudinal and transverse components along the $b$-axis
(figure~\ref{struktura}).

Pseudospin variables
$\frac{\sigma_{q1}}{2},\dots,\frac{\sigma_{q4}}{2}$ describe the changes connected with reorientation of the dipole moments of the base units: $d_{qf} = \mu_f \frac{\sigma_{qf}}{2}$.
Mean values $\langle \frac{\sigma}{2}\rangle = \frac12
(n_a-n_b)$ are connected with the differences in the occupancy of the two possible molecular positions, $n_a$ and $n_b$.

Herein below for convenience we often use the notations $1$, $2$ and $3$ instead of $x$, $y$ and~$z$ for components of vectors and tensors.
The Hamiltonian of a proton subsystem of GPI, which takes into account the short-range and long-range interactions and the applied electric fields $E_1$, $E_2$, $E_3$ along positive directions of the Descartes axes $OX$, $OY$ and $OZ$, consists of the ``seed'' and pseudospin parts. The ``seed'' energy $U_{\text{seed}}$ corresponds to the heavy ion sublattice and does not depend explicitly on the configuration of the proton subsystem. The pseudospin part describes  short-range $\hat H_{\text{short}}$ and long-range $\hat H_{\text{MF}}$ interactions of protons near tetrahedra  HPO$_3$, as well as the effective interaction with the electric fields  $E_1$, $E_2$ and $E_3$. Therefore,
\bea
&& \hat H= N U_{\text{seed}} + \hat H_{\text{short}} + \hat H_{\text{MF}}\,,
\eea
where $N$  is the total number of primitive cells.

The $U_{\text{seed}}$ corresponds to the ``seed'' energy, which includes the elastic, piezoelectric and dielectric parts, expressed in terms of electric fields
 $E_i$ $(i=1, 2, 3)$ and strains  $\varepsilon_i$ and $\varepsilon_j$
$(j=i+3)$. Parameters $c_{ii'}^{E0}(T)$, $c_{i5}^{E0}(T)$, $c_{46}^{E0}(T)$,
 $c_{jj}^{E0}(T)$, $e_{ii'}^0$, $e_{ij}^0$, $\chi_{ii}^{\varepsilon 0}$,
 $\chi_{31}^{\varepsilon 0}$ $(i'=1, 2, 3)$ correspond to the so-called ``seed'' elastic constants,
piezoelectric stresses and dielectric susceptibilities, respectively, $v$ is the volume of a primitive cell:
\begin{align}
 U_{\text{seed}}&= v\Bigg[\frac{1}{2}\sum\limits_{i,i'=1}^3c_{ii'}^{E0}(T)\varepsilon_i \varepsilon_{i'}+
\frac{1}{2}\sum\limits_{j=4}^6c_{jj}^{E0}(T)\varepsilon_j^{2}
 + \sum\limits_{i=1}^3c_{i5}^{E0}(T)\varepsilon_i\varepsilon_5 + c_{46}^{E0}(T)\varepsilon_4\varepsilon_6  \nonumber\\
 &\quad-\sum\limits_{i=1}^3 e_{2i}^0 \varepsilon_i E_2 - e_{25}^0 \varepsilon_5 E_2 -
  e_{14}^0 \varepsilon_4 E_1 - e_{16}^0 \varepsilon_6 E_1-  e_{34}^0 \varepsilon_4 E_3 - e_{36}^0 \varepsilon_6 E_3
  \nonumber\\
 &\quad- \frac{1}{2}  \chi_{11}^{\varepsilon 0}E_1^2 - \frac{1}{2}
\chi_{22}^{\varepsilon 0}E_2^2 -
\frac{1}{2}  \chi_{33}^{\varepsilon 0}E_3^2- \chi_{31}^{\varepsilon 0}E_3E_1\Bigg].
 \end{align}

The Hamiltonian of short-range interactions is
\be
\hat H_{\text{short}} = - 2w \sum\limits_{qq'} \left ( \frac{\sigma_{q1}}{2} \frac{\sigma_{q2}}{2} + \frac{\sigma_{q3}}{2}\frac{\sigma_{q4}}{2} \right)
\left( \delta_{{\vec R}_q{\vec R}_{q'}} + \delta_{{\vec R}_q + {\vec R_{c}},{\vec R}_{q'}} \right). \label{Hshort}
\ee
In (\ref{Hshort}),  $\sigma_{qf}$ is the $z$-component of pseudospin operator that describes the state of the $f$-th bond ($f = 1, 2, 3, 4$), in the  $q$-th cell.
The first Kronecker delta corresponds to the interaction between protons in the chains near the tetrahedra HPO$_{3}$ of type ``I'' (figure~\ref{struktura}), where the second  one near the tetrahedra HPO$_{3}$ of type ``II'', ${\vec R_{c}}$ is the lattice vector along $OZ$-axis. Contributions into the energy of interactions between protons near tetrahedra of different types, as well as the mean values of the pseudospins  $\langle\sigma_{qf}\rangle$, which are related to tetrahedra of different types, are equal.

Parameter $w$, which describes the short-range interactions within chains, is expanded linearly into series over strains $\varepsilon_i$, $\varepsilon_j$:
\be
w = w^{0} + \sum\limits_{i=1}^3 \delta_{i}\varepsilon_i + \sum\limits_{j=4}^6 \delta_{j}\varepsilon_j. \label{w}
\ee
Mean field Hamiltonian $\hat H_{\text{MF}}$  of the long-range dipole-dipole interactions and indirect  (through the lattice vibrations)  interactions between protons,
taking into account that Fourier transforms of interaction constants $J_{ff'} = \sum\limits_{q'} J_{ff'}(qq')$ at $\vec{k}
 =0$ are linearly expanded:
 \begin{eqnarray}
 && J_{ff'} = J^0_{ff'} + \frac{\partial J_{ff'}}{\partial \varepsilon_i}\varepsilon_i
 = J^0_{ff'} + \sum\limits_{i=1}^3\psi_{ff'i}\varepsilon_i + \sum\limits_{j=4}^6\psi_{ff'j}\varepsilon_j\,,
 \end{eqnarray}
can be written as:
\be \hat H_{\text{MF}} = N H^{0} + \hat H_{\text s}\,, \ee
where
\begin{align}
\ H^{0} &=
   \frac18 J^{0}_{11}(\eta_1^2 + \eta_3^2) +\frac18 J^{0}_{22}(\eta_2^2  + \eta_4^2) + \frac14 J_{13}^{0}\eta_1\eta_3
+\frac14 J_{24}^{0}\eta_2\eta_4 +\frac14 J_{12}^{0}(\eta_1\eta_2 + \eta_3\eta_4) +
\frac14 J_{14}^{0}(\eta_1\eta_4 \nonumber\\
&\quad + \eta_2\eta_3)
+ \frac18 \Bigg(\sum\limits_{i=1}^3\psi_{11i}\varepsilon_i+ \sum\limits_{j=4}^6\psi_{11j}\varepsilon_j\Bigg)(\eta_1^2 + \eta_3^2) + \frac18 \Bigg(\sum\limits_{i=1}^3\psi_{22i}\varepsilon_i+ \sum\limits_{j=4}^6\psi_{22j}\varepsilon_j\Bigg)(\eta_2^2  + \eta_4^2) \nonumber\\
& \quad+ \frac14 \Bigg(\sum\limits_{i=1}^3
 \psi_{13i}\varepsilon_i + \sum\limits_{j=4}^6\psi_{13j}\varepsilon_j\Bigg)\eta_1\eta_3 +\frac14 \Bigg(\sum\limits_{i=1}^3
 \psi_{24i}\varepsilon_i + \sum\limits_{j=4}^6\psi_{24j}\varepsilon_j\Bigg)\eta_2\eta_4 \nonumber\\
& \quad+ \frac14  \Bigg(\sum\limits_{i=1}^3  \psi_{12i}\varepsilon_i + \sum\limits_{j=4}^6\psi_{12j}\varepsilon_j\Bigg)(\eta_1\eta_2 + \eta_3\eta_4)+ \frac14 \Bigg( \sum\limits_{i=1}^3  \psi_{14i}\varepsilon_i
+\sum\limits_{j=4}^6\psi_{14j}\varepsilon_j\Bigg)(\eta_1\eta_4 + \eta_2\eta_3), \label{H0}\\
\hat H_{\text s} &= - \sum\limits_q \Bigg( {\cal H}_1
\frac{\sigma_{q1}}{2} + {\cal H}_2 \frac{\sigma_{q2}}{2} + {\cal
H}_3 \frac{\sigma_{q3}}{2} + {\cal H}_4 \frac{\sigma_{q4}}{2}
\Bigg), \label{Hs}
\end{align}
and $\eta_{f}=\langle\sigma_{qf}\rangle$.
In  (\ref{Hs}) the notations are used:
\bea && {\cal H}_1 = \frac{1}{2}J_{11}\eta_1 + \frac{1}{2}J_{12}\eta_2 +
\frac{1}{2}J_{13}\eta_3 + \frac{1}{2}J_{14}\eta_4+
 \mu_{13}^{x}E_1 + \mu_{13}^{y}E_2 + \mu_{13}^{z}E_3\,, \nonumber\\
&& {\cal H}_2 = \frac{1}{2}J_{22}\eta_2 + \frac{1}{2}J_{12}\eta_1 +
\frac{1}{2}J_{24}\eta_4 + \frac{1}{2}J_{14}\eta_3
 - \mu_{24}^{x}E_1 - \mu_{24}^{y}E_2 + \mu_{24}^{z}E_3\,,\nonumber \\
&& {\cal H}_3 = \frac{1}{2}J_{11}\eta_3 + \frac{1}{2}J_{12}\eta_4 +
\frac{1}{2}J_{13}\eta_1+ \frac{1}{2}J_{14}\eta_2
 - \mu_{13}^{x}E_1 + \mu_{13}^{y}E_2 - \mu_{13}^{z}E_3\,,\nonumber\\
&& {\cal H}_4 = \frac{1}{2}J_{22}\eta_4 + \frac{1}{2}J_{12}\eta_3 +
\frac{1}{2}J_{24}\eta_2 + \frac{1}{2}J_{14}\eta_1
  +\mu_{24}^{x}E_1 - \mu_{24}^{y}E_2 - \mu_{24}^{z}E_3.\label{H1234}
\eea
In (\ref{H1234}) $\mu_{13}^{x,y,z}=\mu_{1}^{x,y,z}=\mu_{3}^{x,y,z}$, $\mu_{24}^{x,y,z}=\mu_{2}^{x,y,z}=\mu_{4}^{x,y,z}$  are the effective dipole moments per one pseudospin.

The two-particle cluster approximation is used for calculation of thermodynamic and dielectric characteristics of GPI. In this approximation, thermodynamic potential is given by:
\bea
&& G = N U_{\text{seed}} + NH^0
 - k_{\text B} T  \sum\limits_q \Bigg[ 2\ln \Sp \re^{-\beta \hat H^{(2)}_{q}}
  - \sum\limits_{f=1}^4\ln \Sp  \re^{-\beta \hat H^{(1)}_{qf}} \Bigg], \label{G}
\eea
where $\hat H^{(2)}_{q}$, $\hat H^{(1)}_{qf}$ are two-particle and one-particle Hamiltonians:
\bea
&& \hat H^{(2)}_{q} = - 2w \left( \frac{\sigma_{q1}}{2} \frac{\sigma_{q2}}{2} + \frac{\sigma_{q3}}{2}\frac{\sigma_{q4}}{2}\right)
 - \frac{y_1}{\beta}  \frac{\sigma_{q1}}{2} - \frac{y_2}{\beta} \frac{\sigma_{q2}}{2}  -
\frac{y_3}{\beta}  \frac{\sigma_{q3}}{2} - \frac{y_4}{\beta} \frac{\sigma_{q4}}{2}\, , \label{H2}\\
&&
\hat H^{(1)}_{qf} = - \frac{\bar y_f}{\beta}\frac{\sigma_{qf}}{2}\,. \label{H1}
\eea
Here:
\bea
&& \hspace{-4ex} y_f = \beta (  \Delta_f + {\cal H}_f),  \qquad \bar y_f =  \beta \Delta_f + y_f. \label{yf}
\eea
The symbols $\Delta_f$ are the effective fields created by the neighboring bonds
from outside of the cluster. In the cluster approximation, the fields $\Delta_f$ can be determined from the self-consistency
condition, which states that the mean values of the pseudospins $\langle \sigma_{qf} \rangle$ calculated with the two-particle
and one-particle Gibbs distribution, respectively, should coincide.
That is,
\bea
\frac{\Sp \sigma_{qf} \re^{-\beta \hat H^{(2)}_{q}}}{\Sp \re^{-\beta \hat H^{(2)}_{q}}} =
\frac{\Sp \sigma_{qf} \re^{-\beta \hat H^{(1)}_{qf}}}{\Sp \re^{-\beta \hat H^{(1)}_{qf}}}\,. \label{Sp}
\eea

Hence, based on (\ref{Sp}) taking into account (\ref{H2}) and (\ref{H1}) we obtain
\begin{align}
 \eta_{1,3} &= \frac{1}{D}\left(\sinh n_{1}\pm\sinh  n_{2}+a^{2}\sinh n_{3}\pm a^{2}\sinh  n_{4}+a\sinh  n_{5}+a\sinh  n_{6} \mp a\sinh  n_{7}\pm a\sinh  n_{8}\right)\nonumber\\
 &= \tanh \frac{\bar y_{1,3}}{2}\,,\nonumber\\
 \eta_{2,4} &= \frac{1}{D}\left(\sinh  n_{1}\pm\sinh n_{2}- a^{2}\sinh  n_{3}\mp a^{2}\sinh  n_{4}\mp a\sinh  n_{5}\pm a\sinh  n_{6} +a\sinh  n_{7}+ a\sinh  n_{8}\right) \nonumber\\
 &= \tanh \frac{\bar y_{2,4}}{2}\,, \nonumber\\
   D &= \cosh  n_{1}+\cosh  n_{2} + a^{2}\cosh  n_{3}+ a^{2}\cosh  n_{4}+a\cosh  n_{5}+a\cosh  n_{6}
+ a\cosh  n_{7}+ a\cosh  n_{8}\,,\label{eta}
\end{align}
where
\bea
&&a = \exp\Bigg[-\frac{1}{k_{\text B}T}\Bigg( w^0 + \sum\limits_{i=1}^3\delta_i\varepsilon_i + \sum\limits_{j=4}^6\delta_j\varepsilon_j \Bigg)\Bigg],\nonumber\\
&&n_{1}=\frac{1}{2}(y_1+y_2+y_3+y_4),\qquad  n_{2}=\frac{1}{2}(y_1+y_2-y_3-y_4),\nonumber\\
&&n_{3}=\frac{1}{2}(y_1-y_2+y_3-y_4), \qquad n_{4}=\frac{1}{2}(y_1-y_2-y_3+y_4),\nonumber\\
&&n_{5}=\frac{1}{2}(y_1-y_2+y_3+y_4), \qquad n_{6}=\frac{1}{2}(y_1+y_2+y_3-y_4),\nonumber\\
&&n_{7}=\frac{1}{2}(-y_1+y_2+y_3+y_4),\qquad n_{8}=\frac{1}{2}(y_1+y_2-y_3+y_4).\nonumber
\eea

Taking into consideration (\ref{eta}), we exclude the parameters $\Delta_f$  and write the relations
\bea
&& y_1 = \frac{1}{2} \ln \frac{1 +  \eta_{1}} {1 -  \eta_{1}}+\beta\nu_{11}\eta_{1}+\beta\nu_{12}\eta_{2}+\beta\nu_{13}\eta_{3}+\beta\nu_{14}\eta_{4}+\frac{\beta}{2}(\mu_{13}^{x}E_1 + \mu_{13}^{y}E_2 + \mu_{13}^{z}E_3),\nonumber\\
&& y_2 = \beta\nu_{12} \eta_{1}+\frac{1}{2} \ln \frac{1 +  \eta_{2}} {1 -  \eta_{2}}+\beta\nu_{22}\eta_{2}+\beta\nu_{14}\eta_{3}+\beta\nu_{24}\eta_{4}+\frac{\beta}{2}(- \mu_{24}^{x}E_1 - \mu_{24}^{y}E_2 + \mu_{24}^{z}E_3),\nonumber\\
&& y_3 = \beta\nu_{13} \eta_{1}+\beta\nu_{14}\eta_{2}+\frac{1}{2} \ln \frac{1 +  \eta_{3}} {1 -   \eta_{3}}+\beta\nu_{11}\eta_{3}+\beta\nu_{12}\eta_{4}+\frac{\beta}{2}( - \mu_{13}^{x}E_1 + \mu_{13}^{y}E_2 - \mu_{13}^{z}E_3),\nonumber\\
&& y_4 = \beta\nu_{14} \eta_{1}+\beta\nu_{24}\eta_{2}+\beta\nu_{12}\eta_{3}+\frac{1}{2} \ln \frac{1 +  \eta_{4}} {1 -  \eta_{4}}+\beta\nu_{22}\eta_{4}+\frac{\beta}{2}(\mu_{24}^{x}E_1 - \mu_{24}^{y}E_2 - \mu_{24}^{z}E_3),\nonumber
\eea

where  $\nu_{ff'}=\frac{J_{ff'}}{4}$.

\section{Dielectric characteristics of GPI}

To calculate the dielectric, piezoelectric and elastic characteristics of the GPI, we use the thermodynamic potential per one primitive cell obtained in the two-particle cluster approximation:
  \begin{align}
 g &= \frac{G}{N} = U_{\text{seed}}+H^{0}- 2 \Bigg(w^{0} + \sum\limits_{i=1}^3 \delta_{i}\varepsilon_i +  \sum\limits_{j=4}^6 \delta_{j}\varepsilon_i\Bigg) - \frac{1}{2}k_{\text B}T \sum\limits_{f=1}^4\ln ( 1 - \eta_{f}^{2})  - 2k_{\text B}T \ln D \nonumber \\
   &\quad + 2k_{\text B}T\ln2.
 \end{align}

Minimizing the thermodynamic potential with respect to the strains $\varepsilon_{i}$, $\varepsilon_{j}$, we have obtained equations for the strains:
 \begin{align}
  0 &= c_{l1}^{E0}\varepsilon_1 + c_{l2}^{E0}\varepsilon_2 + c_{l3}^{E0}\varepsilon_3 + c_{l5}^{E0}\varepsilon_5 - e_{2l}^0E_2 -\frac{2\delta_{l}}{\upsilon}+ \frac{2\delta_l}{v D}M_{\varepsilon}
 -  \frac{\psi_{11l}}{8v} (\eta_{1}^{2}+\eta_{3}^{2})-\frac{\psi_{13l}}{4v} \eta_{1}\eta_{3}\nonumber\\
 &\quad-\frac{\psi_{22l}}{8v} (\eta_{2}^{2}+\eta_{4}^{2})-   \frac{\psi_{24l}}{4v} \eta_{2}\eta_{4}
- \frac{\psi_{12l}}{4v} (\eta_{1}\eta_{2}+\eta_{3}\eta_{4})- \frac{\psi_{14l}}{4v} (\eta_{1}\eta_{4}+\eta_{2}\eta_{3}),\qquad(l=1,2,3,5) \nonumber\\
  0 &= c_{44}^{E0}\varepsilon_4+c_{46}^{E0}\varepsilon_6
 - e_{14}^0 E_1 - e_{34}^0 E_3-\frac{2\delta_{4}}{\upsilon}+ \frac{2\delta_4}{v D}M_{\varepsilon}
 -  \frac{\psi_{114}}{8v} (\eta_{1}^{2}+\eta_{3}^{2})-  \frac{\psi_{134}}{4v} \eta_{1}\eta_{3}\nonumber\\
 &\quad-\frac{\psi_{224}}{8v} (\eta_{2}^{2}+\eta_{4}^{2})-  \frac{\psi_{244}}{4v} \eta_{2}\eta_{4}
- \frac{\psi_{124}}{4v} (\eta_{1}\eta_{2}+\eta_{3}\eta_{4})- \frac{\psi_{144}}{4v} (\eta_{1}\eta_{4}+\eta_{2}\eta_{3}),\nonumber\\
  0 &= c_{46}^{E0}\varepsilon_4 +
c_{66}^{E0}\varepsilon_6 - e_{16}^0 E_1 - e_{36}^0 E_3 -\frac{2\delta_{6}}{\upsilon}+ \frac{2\delta_6}{v D}M_{\varepsilon}-  \frac{\psi_{116}}{8v} (\eta_{1}^{2}+\eta_{3}^{2})-\frac{\psi_{136}}{4v} \eta_{1}\eta_{3}\nonumber\\
&\quad-\frac{\psi_{226}}{8v} (\eta_{2}^{2}+\eta_{4}^{2})-    \frac{\psi_{246}}{4v} \eta_{2}\eta_{4} - \frac{\psi_{126}}{4v} (\eta_{1}\eta_{2}+\eta_{3}\eta_{4})- \frac{\psi_{146}}{4v} (\eta_{1}\eta_{4}+\eta_{2}\eta_{3}),\label{sigma}
 \end{align}
where
\bea
&& M_{\varepsilon}=2a^{2}\cosh n_{3}+2a^{2}\cosh n_{4}
+a \cosh n_{5}+a\cosh n_{6}+a \cosh n_{7}+a\cosh n_{8}.\nonumber
 \eea

Differentiating the thermodynamic potential over the fields $E_{i}$ we get the expressions for polarizations~$P_{i}$
 \bea
&& P_1 = e_{14}^0\varepsilon_4 + e_{16}^0\varepsilon_6 +\chi_{11}^{\varepsilon 0}E_1+ \chi_{31}^{\varepsilon 0}E_3+ \frac{1}{2v}[\mu_{13}^{x}(\eta_{1}-\eta_{3})-\mu_{24}^{x}(\eta_{2}-\eta_{4})], \nonumber \\
 && P_2 = e_{21}^0\varepsilon_1 + e_{22}^0\varepsilon_2 +
 e_{23}^0\varepsilon_3 + e_{25}^0\varepsilon_5  + \chi_{22}^{\varepsilon 0}E_2  + \frac{1}{2v}[\mu_{13}^{y}(\eta_{1}+\eta_{3})-\mu_{24}^{y}(\eta_{2}+\eta_{4})],   \nonumber\\
 && P_3 =  e_{34}^0\varepsilon_4 + e_{66}^0\varepsilon_6+  \chi_{33}^{\varepsilon 0}E_3 +  \chi_{31}^{\varepsilon 0}E_1+ \frac{1}{2v}[\mu_{13}^{z}(\eta_{1}-\eta_{3})+\mu_{24}^{z}(\eta_{2}-\eta_{4})].\label{P}
 \eea
Diagonal components of the static isothermic dielectric susceptibilities of mechanically clamped crystal
GPI are given by:
\bea
&&\chi_{11}^{\varepsilon}=\chi_{11}^{\varepsilon 0}+
\frac{1}{2\upsilon\Delta}[\mu_{13}^{x}
(\Delta_{1}^{\chi x}- \Delta_{3}^{\chi x})-
\mu_{24}^{x}
(\Delta_{2}^{\chi x}- \Delta_{4}^{\chi x})], \label{X11} \\
&&\chi_{22}^{\varepsilon}=\chi_{22}^{\varepsilon 0}+
\frac{1}{2\upsilon\Delta}[\mu_{13}^{y}
(\Delta_{1}^{\chi y}+ \Delta_{3}^{\chi y})-
\mu_{24}^{y}
(\Delta_{2}^{\chi y}+ \Delta_{4}^{\chi y})], \label{X22}
\\
&&\chi_{33}^{\varepsilon}=\chi_{33}^{\varepsilon 0}+
\frac{1}{2\upsilon\Delta}[\mu_{13}^{z}
(\Delta_{1}^{\chi z}- \Delta_{3}^{\chi z})+
\mu_{24}^{c}
(\Delta_{2}^{\chi z}- \Delta_{4}^{\chi z})]. \label{X33}
\eea
Here, the ratio
\bea
\frac{\Delta_{f}^{\chi \alpha}}{\Delta}=\left( \frac{\partial \eta_f}{\partial
 E_{\alpha}}\right)_{\varepsilon_l} \nonumber
\eea
  has the meaning of the local pseudospin susceptibility, which describes the reaction of the  $f$-th order parameter to the external electric field $E_{\alpha}$ at constant strains. Explicit expressions for quantities introduced here are given in the appendix [formulae~(\ref{A1}) and~(\ref{A2})].

Based on (\ref{sigma}), we have obtained expressions for isothermic coefficients of piezoelectric stress  $e_{2j}$ of GPI:
\bea
 && e_{2l}^{} = \left( \frac{\partial P_2}{\partial
 \varepsilon_l}\right)_{E_2} = e_{2l}^0+ \frac{\mu_{13}^{y}}{2v\Delta}(\Delta_{1l}^{e}+\Delta_{3l}^{e})
-\frac{\mu_{24}^{y}}{2v\Delta}(\Delta_{2l}^{e}+\Delta_{4l}^{e}). \label{e2l}
\eea
Here, the  ratio
\bea
\frac{\Delta_{fl}^{e}}{\Delta}=\left( \frac{\partial \eta_f}{\partial
 \varepsilon_l}\right)_{E_{2}} \nonumber
\eea
describes the reaction of the $f$-th order parameter on the strain $\varepsilon_l$ at constant external fields [see the appendix, formula~(\ref{A3})].

\section{Comparison  with the experimental data}

To calculate the temperature and field dependences of
dielectric and piezoelectric characteristics
of GPI, we have to determine the values of the following parameters:
\begin{itemize}
\item parameter of short-range interactions $w^{0}$;

\item parameters of long-range interactions $\nu_{f}^{0\pm}$ ($f=1,2,3$);

\item deformational potentials $\delta_{i}$,   $\psi_{fi}^{\pm}$ ($f=1,2,3$; $i=1,\ldots,6$);

\item effective dipole moments
$\mu_{13}^{x}$; $\mu_{24}^{x}$; $\mu_{13}^{y}$; $\mu_{24}^{y}$; $\mu_{13}^{z}$; $\mu_{24}^{z}$;

\item ``seed'' dielectric susceptibilities $\chi_{ii}^{\varepsilon 0}$,
 $\chi_{31}^{\varepsilon 0}$  $(i=1, 2, 3)$;

\item ``seed'' coefficients of piezoelectric stress $e_{2i}^0$, $e_{25}^0$, $e_{14}^0$, $e_{16}^0$, $e_{34}^0$, $e_{36}^0$;

\item ``seed''  elastic constants $c_{ii'}^{E0}$, $c_{jj}^{E0}$, $c_{i5}^{E0}$, $c_{46}^{E0}$  ($i=1, 2, 3$; $i'=1, 2, 3$; $j=4, 5, 6$).
\end{itemize}

To determine the above listed parameters, we use the measured temperature dependences  for the set of physical
characteristics of GPI,  namely  $P_{\text s}(T)$ \cite{nay1}, $\varepsilon_{11}^\sigma$,  $\varepsilon_{33}^\sigma$ \cite{dac}, $d_{21}$, $d_{23}$ \cite{wie}, as well as the dependence of phase transition temperature $T_{\text c}(p)$ \cite{yas} on hydrostatic pressure.

The volume of primitive cell of GPI is the  $\upsilon_{\text H}$ = 0.601$\cdot 10^{-21}$~cm$^3$ \cite{Taniguchi2003}.

Numerical analysis shows that thermodynamic characteristics depend on the two linear combinations of long-range interactions
$\nu^{0+}=\nu_1^{0+}+2\nu_2^{0+}+\nu_3^{0+}$ and $\nu^{0-}=\nu_1^{0-}+2\nu_2^{0-}+\nu_3^{0-}$ and practically do not depend  (deviation $<0.1$\%) on separate values of the $\nu_{f}^{0\pm}$ at given $\nu^{0+}$ and $\nu^{0-}$. The optimal values of these combinations are  $\nu^{0+}/k_{\text B}=10.57$~K, $\nu^{0-}/k_{\text B}=-0.8$~K; as concrete values of the  $\nu_{f}^{0\pm}$ we use $\tilde \nu_1^{0+}=\tilde \nu_2^{0+}=\tilde \nu_3^{0+}=2.643$~K, $\tilde \nu_1^{0-}=\tilde \nu_2^{0-}=\tilde \nu_3^{0-}=0.2$~K, where $\tilde \nu_{f}^{0\pm}=\nu_{f}^{0\pm}/k_{\text B}$.

Since the phase transition in the GPI is of the second order, from the condition of nullification of the inverse longitudinal dielectric susceptibility (\ref{X22}) we can obtain the equation $\Delta(T_{\text c})= 0$ for phase transition temperature. This equation connects the parameter of short-range interactions $w^0$  with the parameters of long-range interactions $\nu_1^{0+}$, $\nu_2^{0+}$ and $\nu_3^{0+}$.
From this equation at $\vec{E}=0$ and at the given $\nu_1^{0+}$, $\nu_2^{0+}$, $\nu_3^{0+}$ and other parameter values, we obtain the value of the short-range parameter $w^0$.
Its optimal value is $w^0=820$~K.
The optimal values of deformational potentials $\delta_{j}$, which are coefficients of linear expansion of the parameter  $w^0$ over the strains $\varepsilon_{j}$ [see~(\ref{w})], are as follows:
$\tilde\delta_{1}=500$~K,  $\tilde\delta_{2}=600$~K, $\tilde\delta_{3}=500$~K, $\tilde\delta_{4}=150$~K, $\tilde\delta_{5}=100$~K, $\tilde\delta_{6}=150$~K; $\tilde\delta_{i}={\delta_{i}}/{k_{\text B}}$.

For parameters $\psi_{fi}^{\pm}$, similarly to the $\nu_{f}^{0\pm}$, the 6 linear combinations  $\psi_{i}^{+}=\psi_{1i}^{+}+2\psi_{2i}^{+}+\psi_{3i}^{+}$ and 6 combinations $\psi_{i}^{-}=\psi_{1i}^{-}+2\psi_{2i}^{-}+\psi_{3i}^{-}$ are important. Thermodynamic characteristics practically do not depend (deviation $<0.1$\%) on separate values of the  $\psi_{fi}^{\pm}$ at given $\psi_{i}^{+}$ and $\psi_{i}^{-}$.
The optimal values of the $\psi_{fi}^{\pm}$, are as follows:
$\tilde\psi_{f1}^{+} = 87.9$~K,  $\tilde\psi_{f2}^{+}= 237.0$~K,  $\tilde\psi_{f3}^{+}= 103.8$~K,
$\tilde\psi_{f4}^{+} = 149.1$~K,  $\tilde\psi_{f5}^{+} = 21.3$~K,  $\tilde\psi_{f6}^{+} = 143.8$~K,  $\tilde\psi_{fi}^{-}=0$~K, where
 $\tilde\psi_{fi}^{\pm} =\psi_{fi}^{\pm}/{k_{\text B}}$.

Effective dipole moments in the paraelectric phase are equal to $\vec\mu_{13} = (0.4, 4.02, 4.3)\cdot 10^{-18}$~esu$\cdot$cm,  $\vec\mu_{24} = (-2.3, -3.0, 2.2)\cdot 10^{-18}$~esu$\cdot$cm. In the ferroelectric phase, the $y$-component of the first dipole moment is  $\mu_{13\text{ferro}}^{y}=3.82\cdot 10^{-18}$~esu$\cdot$cm;
X-ray investigation \cite{Shikanai2002} determined the coordinates of atoms in the primitive cell of GPI. The calculated displacements of the protons, which we marked as 1 and 2, relative to the centers of hydrogen bonds in ferroelectric phase are equal to $\Delta \vec r_1=(-0.016, -0.495, -0.160)$~{\AA}, $\Delta \vec r_2=(0.389, 0.383, -0.147)$~{\AA}. The obtained dipole moments are not proportional to the corresponding proton displacements. This means that in addition to the proton displacements, the phosphite and glycine groups also take part in forming the effective dipole moments.

For the ``seed'' coefficients of piezoelectric stress, dielectric susceptibilities and elastic constants, the following values are obtained:
\begin{eqnarray}
&e_{21}^0 = e_{22}^0 = e_{23}^0 = e_{25}^0 = e_{14}^0 = e_{16}^0 = e_{34}^0 = e_{36}^0 = 0.0~\frac{\text{esu}}{\text{cm}^2};
~~\chi_{11}^{\varepsilon 0} = 0.1,~~ \chi_{22}^{\varepsilon 0}= 0.403, ~~\chi_{33}^{\varepsilon 0} = 0.5,&\nonumber\\
& \chi_{31}^{\varepsilon 0} = 0.0; ~~ c_{11}^{0E} = 26.91\cdot10^{10}~\frac{\text{dyn}}{\text{cm}^2}\,,~~
c_{12}^{E0} = 14.5 \cdot 10^{10}~\frac{\text{dyn}}{\text{cm}^2}\,,~~
c_{13}^{E0} = 11.64 \cdot10^{10}~\frac{\text{dyn}}{\text{cm}^2}\,,&\nonumber\\
&c_{15}^{E0} = 3.91  \cdot10^{10} ~\frac{\text{dyn}}{\text{cm}^2}\,,~~
c_{22}^{E0} = [64.99- 0.04(T-T_{\text c})] \cdot10^{10} ~\frac{\text{dyn}}{\text{cm}^2}\,,~~
c_{23}^{E0} = 20.38\cdot10^{10} ~\frac{\text{dyn}}{\text{cm}^2}\,,&\nonumber\\
&c_{25}^{E0} = 5.64  \cdot10^{10} ~\frac{\text{dyn}}{\text{cm}^2}\,,~~
c_{33}^{E0} = 24.41\cdot10^{10} ~\frac{\text{dyn}}{\text{cm}^2}\,,~~
c_{35}^{E0} = -2.84  \cdot10^{10} ~\frac{\text{dyn}}{\text{cm}^2}\,,~~
c_{55}^{E0} = 8.54 \cdot 10^{10}~\frac{\text{dyn}}{\text{cm}^2}\,,&\nonumber\\
&c_{44}^{E0} = 15.31 \cdot10^{10}~\frac{\text{dyn}}{\text{cm}^2}\,,~~
c_{46}^{E0} = -1.1 \cdot 10^{10} ~\frac{\text{dyn}}{\text{cm}^2}\,,~~
c_{66}^{E0} = 11.88 \cdot 10^{10}~\frac{\text{dyn}}{\text{cm}^2}.&\nonumber
\end{eqnarray}

In \cite{Stasyuk2004}, the phase transition temperature of the GPI crystal was $T_{\text c}=222$~K.
Explaining the experimental data \cite{Stasyuk2004} we suppose that all interactions in this crystal are proportional to the interactions in the crystal with  $T_{\text c}=225$~K. Thus, $w^0(222~{\text K})=kw^0(225~{\text K})$, $\nu_{f}^{0\pm}(222~{\text K})=k\nu_{f}^{0\pm}(225~{\text K})$, $\delta_{i}(222~{\text K})=k\delta_{i}(225~{\text K})$, $\psi_{fi}^{\pm}(222~{\text K})=k\psi_{fi}^{\pm}(225~{\text K})$, where $k=0.987\approx222/225$. Besides, the $y$-components of the dipole moments are the same in paraelectric and ferroelectric phases, that is $\mu_{13\text{ferro}}^{y}=\mu_{13\text{para}}^{y}=3.82\cdot 10^{-18}$~esu$\cdot$cm; and $z$-component $\mu_{13}^{z}=4.2\cdot 10^{-18}$~esu$\cdot$cm. All other parameters are taken the same as for the crystal with  $T_{\text c}=225$~K.

Now, let us look at the results obtained in this paper  for temperature and field dependences of physical characteristics of the GPI crystal at different values of strength of the electric fields $E_1$ and~$E_3$.
Numerical calculations of dielectric characteristics of the GPI are carried out for the strength of the fields from  0 up to $\pm 4 $~MV/m.

\begin{figure}[!b]
\centering
\begin{minipage}{0.49\textwidth}
\begin{center}
\includegraphics[scale=0.82]{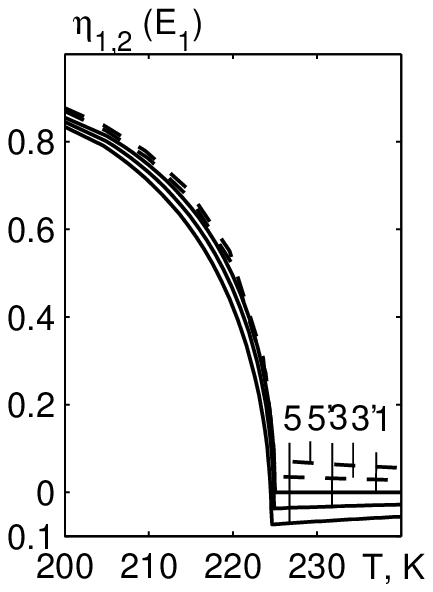}\includegraphics[scale=0.82]{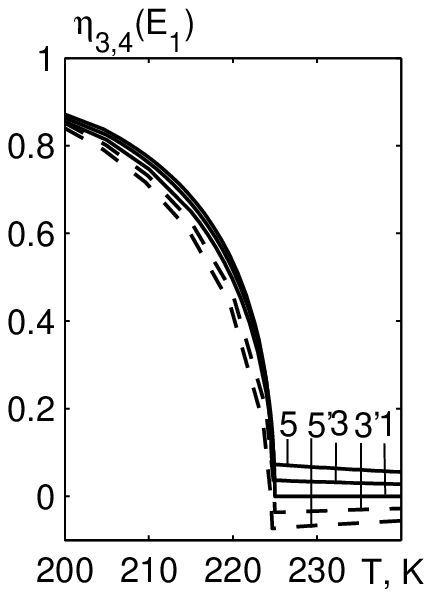}
\caption{The temperature dependences of the order parameters $\eta_{f}$ of the GPI crystal at different values of the electric field $E_1$~(MV/m): 0.0~---~1; 2.0~---~3; $-2.0$~---~3'; 4.0~---~5; $-4.0$~---~5'.}\label{eta1_E1}
\end{center}
\end{minipage}
\begin{minipage}{0.49\textwidth}
\begin{center}
\includegraphics[scale=0.82]{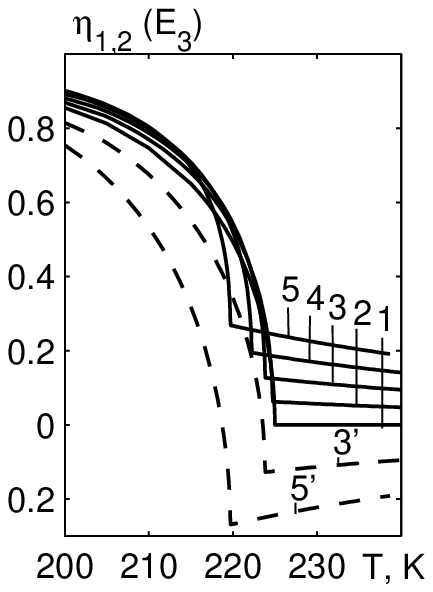}\includegraphics[scale=0.82]{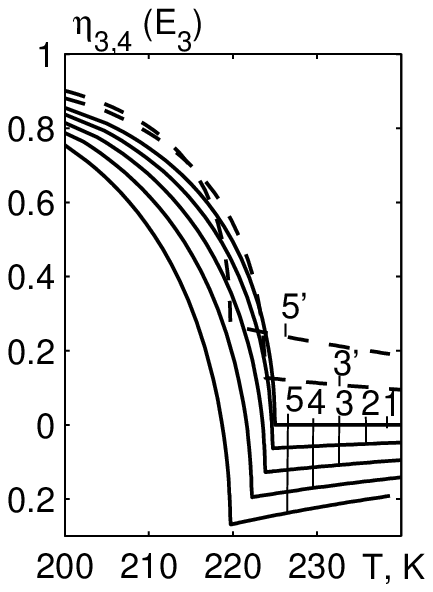}
\caption{The temperature dependences of the order parameters $\eta_{f}$ of the GPI crystal at different values of the electric field  $E_{3}$~(MV/m): 0.0~---~1; 1.0~---~2; 2.0~---~3; 3.0~---~4; 4.0~---~5; $-2.0$~---~3'; $-4.0$~---~5'.}\label{eta1_E3}
\end{center}
\end{minipage}
\end{figure}

Temperature dependences of the order parameters at different values of the fields $E_1$ or $E_3$ are presented in figures~\ref{eta1_E1} and~\ref{eta1_E3}. At zero fields, the mean values of pseudospins are $\eta_1=\eta_3$, $\eta_2=\eta_4$ in the
ferroelectric phase, and $\eta_1=\eta_2=\eta_3=\eta_4=0$ in the paraelectric phase.

The electric field $E_1>0$ slightly splits  the mean values of pseudospins in the
ferroelectric phase, and fairly strongly in the paraelectric phase. In the paraelectric phase, $\eta_1=\eta_2<0$, $\eta_3=\eta_4>0$. An increase of the field  $E_1$ leads to a decrease  of the  $\eta_1$, $\eta_2$ and to an increase of $\eta_3$, $\eta_4$ parameters. In the case of  $E_1<0$,  in the paraelectric phase $\eta_1=\eta_2>0$, $\eta_3=\eta_4<0$.

Applying the electric field $E_3>0$ also leads to a splitting of the mean values of pseudospins, but much stronger than in the case of the field $E_1$. Here,
$\eta_1=\eta_2>0$, $\eta_2=\eta_4<0$ in the paraelectric phase. An increase of the field $E_3$ in the ferroelectric phase leads to an increase of $\eta_1$, $\eta_2$ and to a decrease of $\eta_3$, $\eta_4$ parameters. At $E_3<0$,  in the paraelectric phase $\eta_1=\eta_2<0$, $\eta_2=\eta_4>0$.

\begin{figure}[!t]
\centering
\begin{minipage}{0.49\textwidth}
\begin{center}
\includegraphics[scale=0.85]{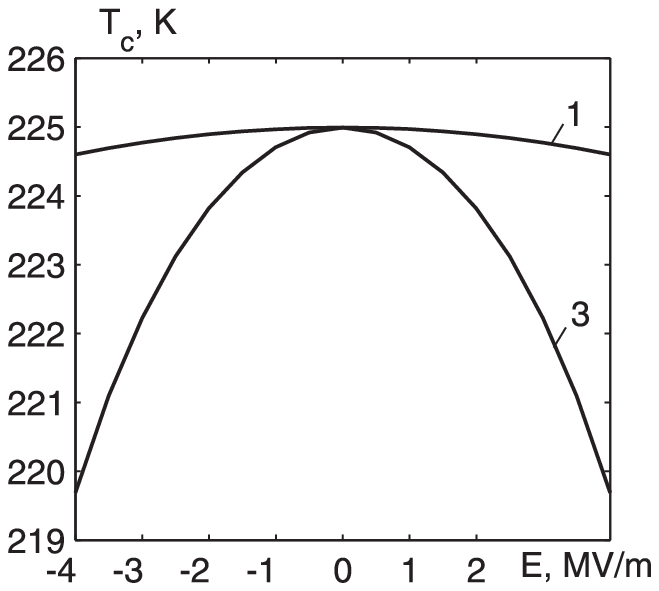}
\end{center}
\end{minipage}
\begin{minipage}{0.49\textwidth}
\begin{center}
\includegraphics[scale=0.85]{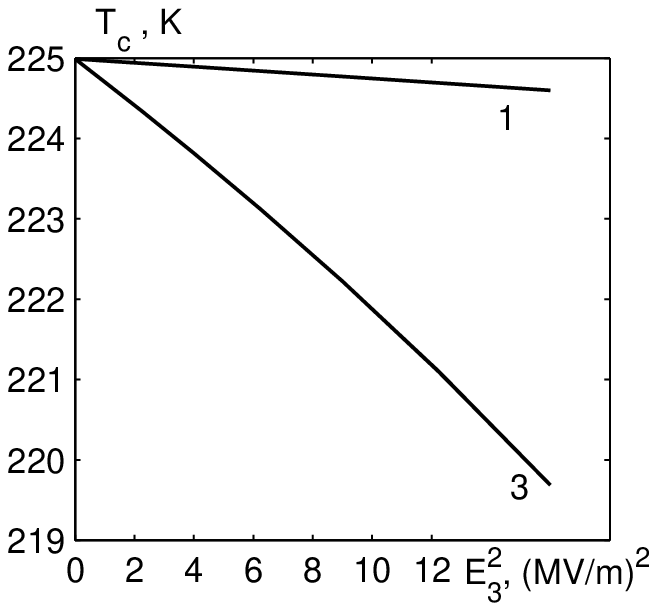}
\end{center}
\end{minipage}
\begin{minipage}{0.49\textwidth}
\begin{center}
\caption{The dependences of the phase transition temperature $T_{\text c}$ of GPI crystal on the electric fields $E_1$~(1) and $E_3$~(3). } \label{Tc_E13}
\end{center}
\end{minipage}
\begin{minipage}{0.49\textwidth}
\begin{center}
\caption{The dependences of the phase transition temperature $T_{\text c}$ of GPI crystal on the squares of the electric fields $E_1$~(1) and $E_3$~(3).}\label{Tc_E3E1}
\end{center}
\end{minipage}
\end{figure}

The dependences of the phase transition temperature $T_{\text c}$ of GPI crystal on the electric fields $E_1$ and $E_3$, and on the squares of these fields are presented in figures~\ref{Tc_E13} and~\ref{Tc_E3E1}, respectively. With an increase of the fields  $E_1$ and $E_3$, the phase transition temperatures $T_{\text c}$ decrease, especially for the field $E_3$.

\begin{figure}[!b]
\begin{center}
\includegraphics[scale=0.85]{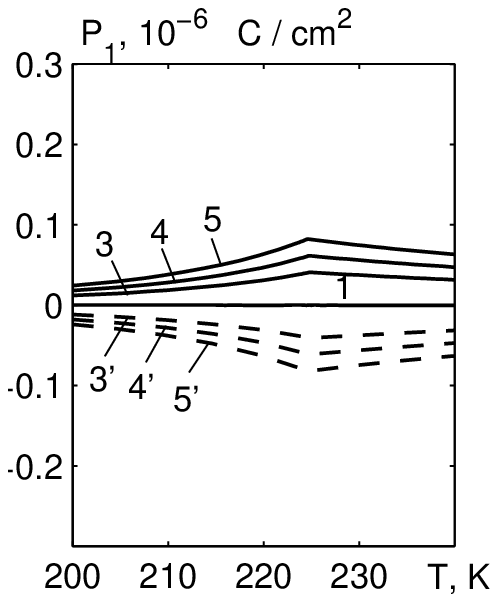}\includegraphics[scale=0.85]{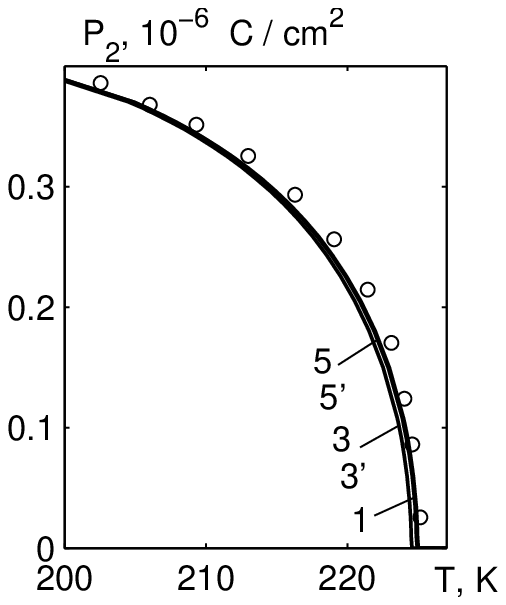} \includegraphics[scale=0.85]{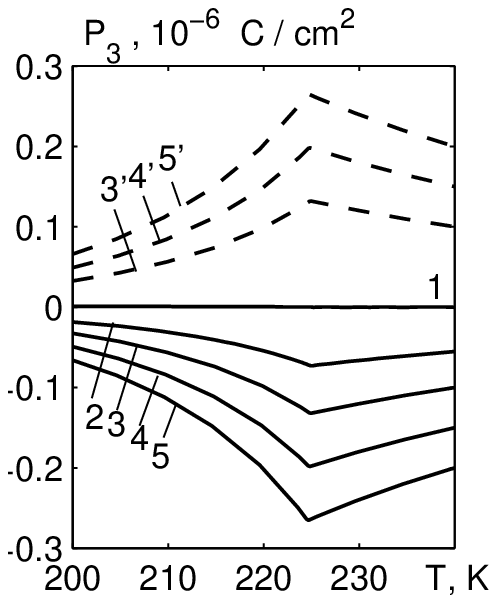}
\end{center}
\caption{ The temperature dependences of the components of polarization $P_{1}$, $P_{2}$, $P_{3}$ of GPI crystal at different values of the field $E_1$~(MV/m): 0.0~---~1; 1.0~---~2; 2.0~---~3; 3.0~---~4; 4.0~---~5; $-$2.0~---~3'; $-$3.0~---~4'; $-$4.0~---~5'; $\circ$ are the experimental data \cite{nay1}.} \label{P1_E1}
\end{figure}
\begin{figure}[!t]
\begin{center}
\includegraphics[scale=0.85]{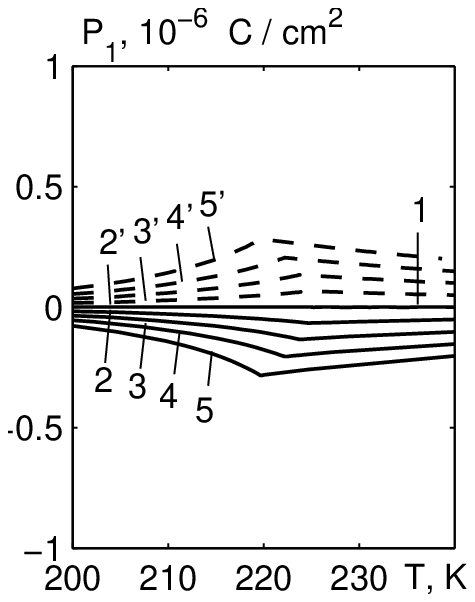}\includegraphics[scale=0.85]{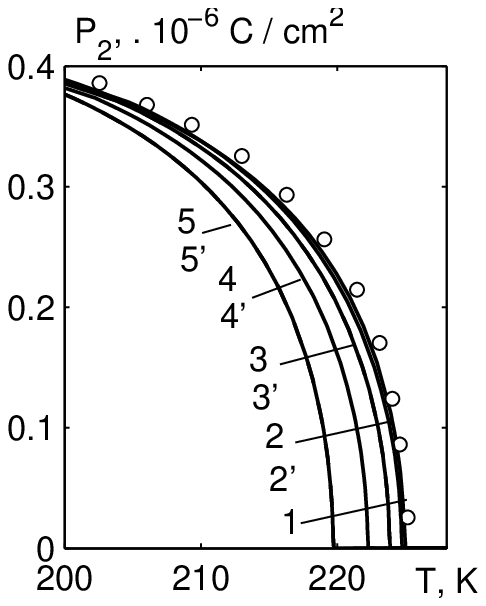} \includegraphics[scale=0.85]{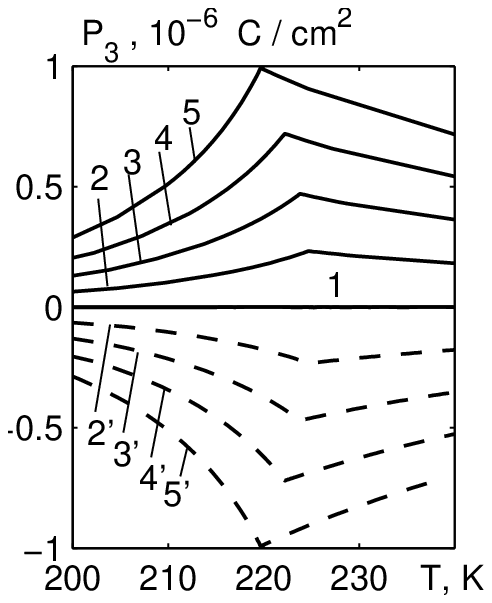}
\end{center}
\caption{The temperature dependences of the components of polarization $P_{1}$, $P_{2}$, $P_{3}$ of GPI crystal at different values of the field $E_3$~(MV/m): 0.0~---~1; 1.0~---~2; 2.0~---~3; 3.0~---~4; 4.0~---~5; $-$1.0 ---2';  $-$2.0~---~3'; $-$3.0~---~4'; $-$4.0~---~5'; $\circ$  are the experimental data \cite{nay1}. } \label{P1_E3}
\end{figure}
It is shown that the dependences $T_{\text c}(E_{1,3})$ are close to quadratic in the fields (see \cite{Stasyuk2004}), and at the fields up to 4~MV/m, they can be written as:
\bea  T_{\text c}(E_{1})=T_{\text c}-k_{1}^{T}E_{1}^{2}, \qquad
T_{\text c}(E_{3})=T_{\text c}-k_{3}^{T}E_{3}^{2}, \nonumber
 \eea
where $k_{1}^{T}=0.025$~Km$^2$/MV$^2$,
$k_{3}^{T}=0.3325$~Km$^2$/MV$^2$.

In figure~\ref{P1_E1} there are presented the  temperature dependences of the components of polarization $P_i$  of GPI crystal at different values of the field $E_1$, and in figure~\ref{P1_E3} --- at different values of the field  $E_3$.

With an increase of strength of the electric field  $E_1$, the spontaneous polarization $P_2$ slightly decreases, but polarization $P_1$  induced by the field  increases. Polarization $P_3$ induced by the field $E_1$ is negative, and in magnitude it is three times larger than the  $P_1$. At the field $E_1<0$, the sign of polarizations $P_1$ and $P_3$ is opposite, and  the magnitude of the $P_2$ also decreases.

\begin{figure}[!t]
\begin{center}
\includegraphics[scale=0.86]{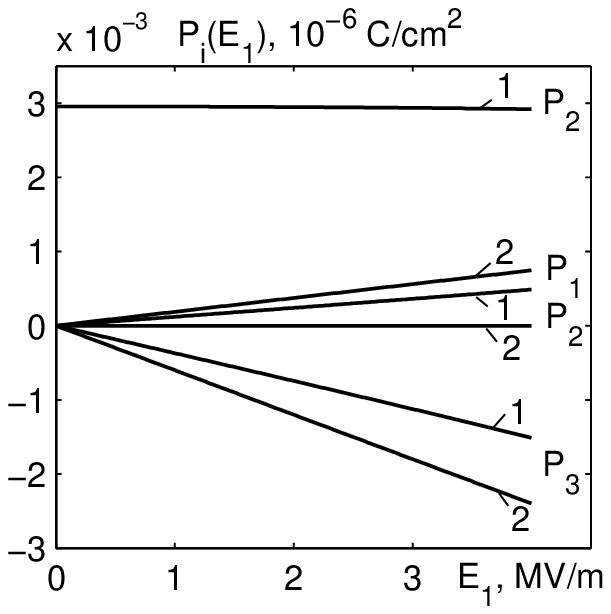}~~\includegraphics[scale=0.86]{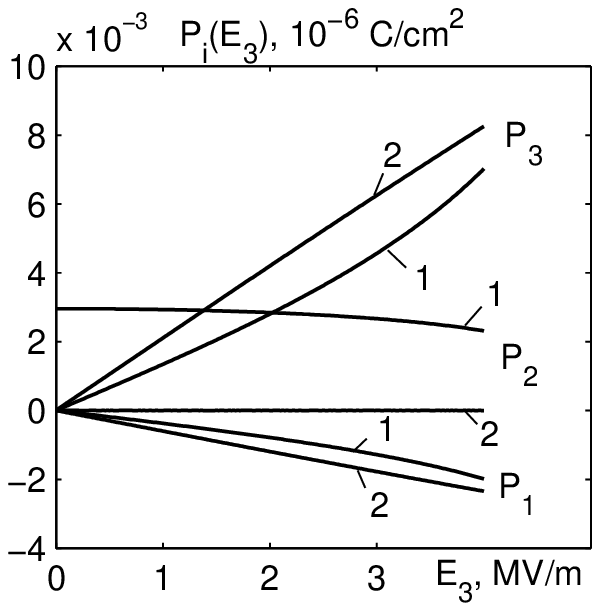}
\end{center}
\caption{ The dependences of polarizations $P_{1}$, $P_{2}$, $P_{3}$ of GPI crystal on the fields  $E_{1}$ and $E_{3}$ at different temperatures  $T$~(K): 215~---~1; 230~---~2. } \label{Pi_TE1}
\end{figure}

However, an increase of the field $E_3$ leads  to a decrease of spontaneous polarization $P_2$ and to an increase of the polarization $P_3$; besides, the $P_3(E_{3})$ increases more appreciably than in the case of $P_1(E_{1})$.
The temperature dependence of the negative polarization $P_1(E_{3})$ induced by the field  $E_3$  is
analogous to the $P_3(E_{1})$ and the value of the $P_1(E_{3})$ is almost equal to the value of the
$P_3(E_{1})$. It is necessary to note that the effect of the field $E_3<0$ on the components of polarization is qualitatively similar to the effect of the field  $E_1>0$ on them.
The dependences of polarizations $P_{1}$, $P_{2}$, $P_{3}$ of GPI crystal on the fields  $E_{1}$ and $E_{3}$ at different temperatures  $T$ are presented in  figure~\ref{Pi_TE1}.

\begin{figure}[!t]
\begin{center}
\includegraphics[scale=0.84]{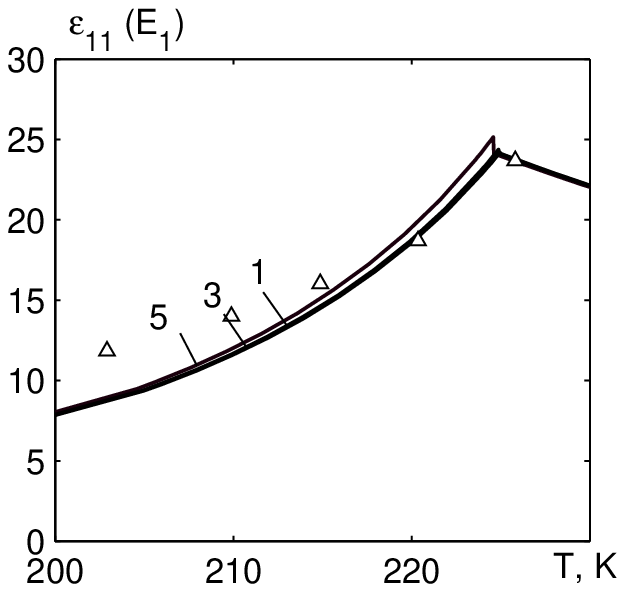}~~\includegraphics[scale=0.84]{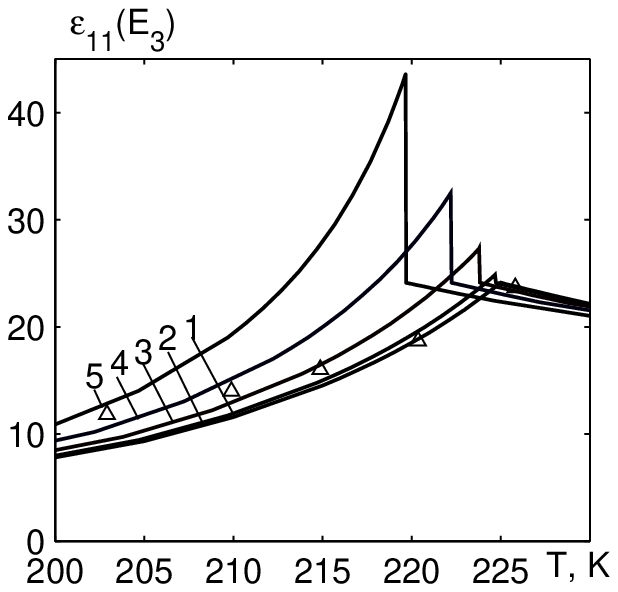}
\end{center}
\caption{ The temperature dependences of the  static dielectric permittivity $\varepsilon_{11}$ of GPI crystal at different values of the fields   $E_1$ and $E_3$~(MV/m): 0.0~---~1; 1.0~---~2; 2.0~---~3; 3.0~---~4; 4.0~---~5;  $\vartriangle$  are the experimental data  \cite{dac}.} \label{eps11_E12}
\end{figure}

\begin{figure}[!t]
\begin{center}
\includegraphics[scale=0.84]{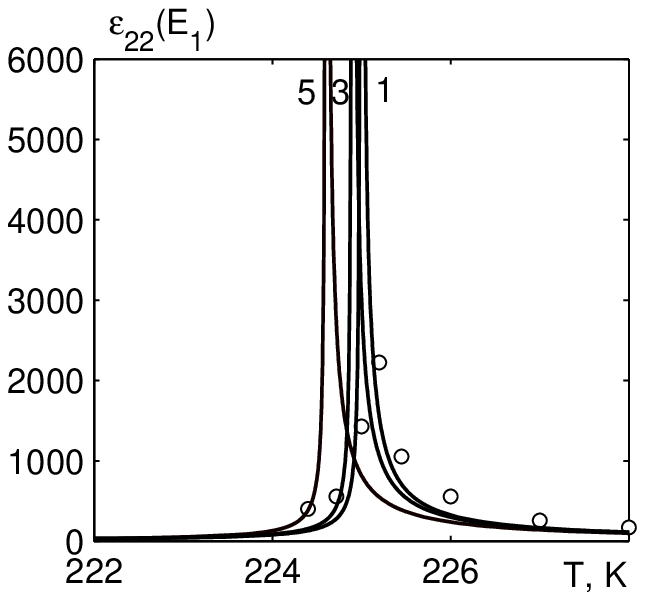}~~~~\includegraphics[scale=0.84]{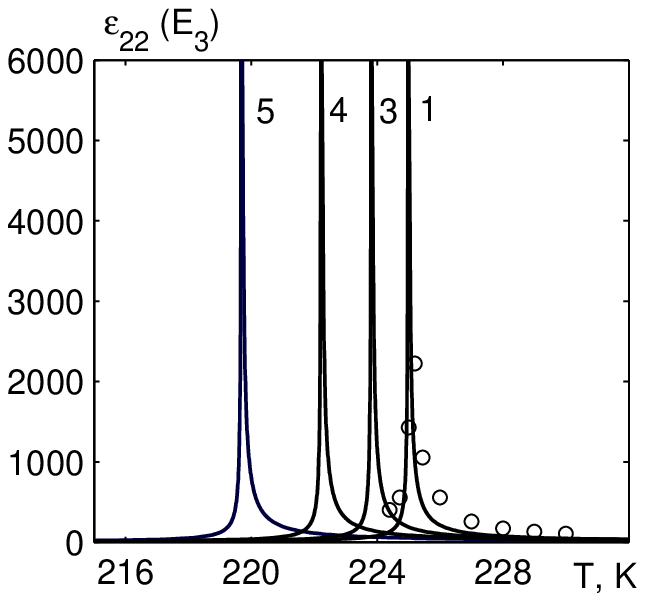}
\end{center}
\caption{ The temperature dependences of the  static dielectric permittivity $\varepsilon_{22}$ of GPI crystal at different values of the fields   $E_1$ and $E_3$~(MV/m): 0.0~---~1; 1.0~---~2; 2.0~---~3; 3.0~---~4; 4.0~---~5; $\circ$ are the experimental data   \cite{nay2}. } \label{eps22_E1}
\end{figure}

\begin{figure}[!t]
\begin{center}
\includegraphics[scale=0.8]{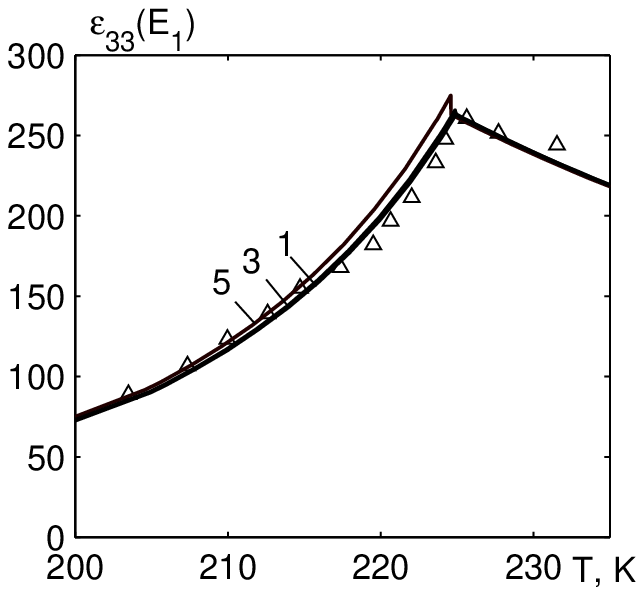}~~\includegraphics[scale=0.8]{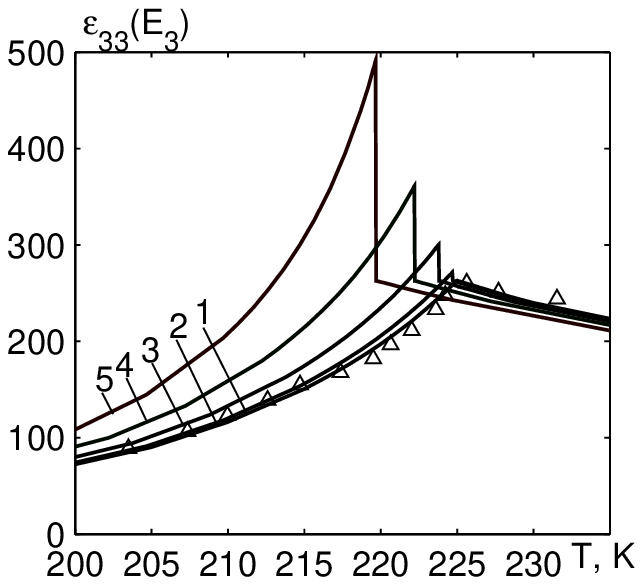}
\end{center}
\caption{ The temperature dependences of the  static dielectric permittivity $\varepsilon_{33}$ of GPI crystal at different values of the fields   $E_1$ and $E_3$~(MV/m): 0.0~---~1; 1.0~---~2; 2.0~---~3; 3.0~---~4; 4.0~---~5; $\vartriangle$  are the experimental data    \cite{dac}.} \label{eps33_E12}
\end{figure}

Changes in the temperature dependences of the  components of static dielectric permittivities
 $\varepsilon_{ii}=1+4\piup \chi_{ii}$ of GPI crystal under the action of transverse electric fields  $E_1$ and $E_3$ are shown in  figures~\ref{eps11_E12}--\ref{eps33_E12}.

Values of the permittivities
$\varepsilon_{11}(E_1)$, $\varepsilon_{33}(E_1)$ slightly increase in the ferroelectric phase and slightly decrease in the paraelectric phase. The action of the field $E_3$ is much stronger.
The temperature dependences of the  $\varepsilon_{11}(E_3)$ and $\varepsilon_{33}(E_3)$ have jumps at the phase transition point, which rise with an increase of the field  $E_3$ and shift to the lower temperatures. Changes in signs of the fields do not influence the values of  permittivities.


\begin{figure}[!t]
\begin{center}
\includegraphics[scale=0.8]{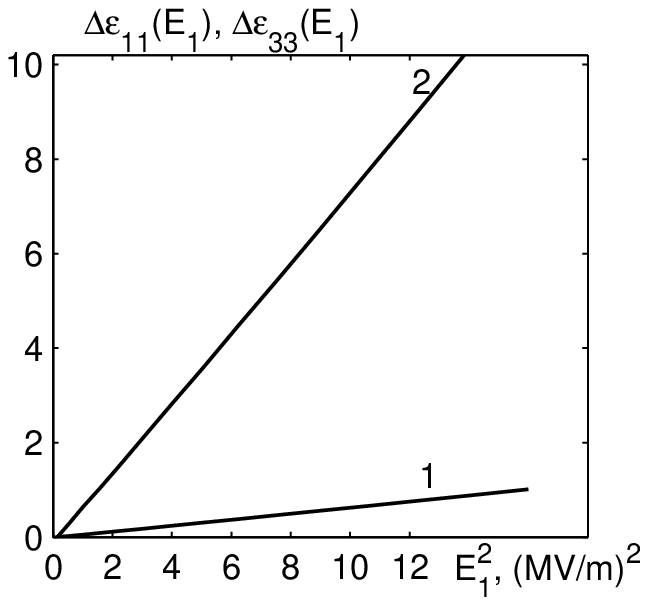}~~\includegraphics[scale=0.8]{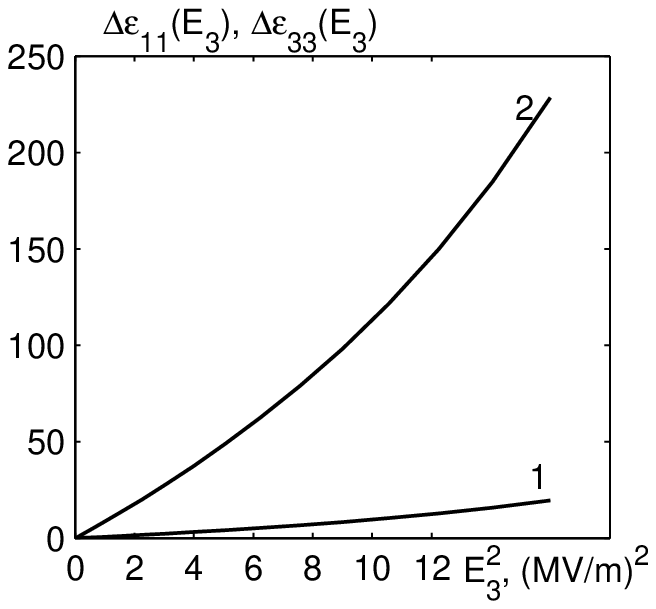}
 \end{center}
 \vspace{-2mm}
\caption[]{The dependences of the jumps of the permittivities $\varepsilon_{11}(1)$ and $\varepsilon_{33}(2)$  of GPI crystal on the squares of the electric fields $E_1$ and $E_3$. } \label{Deltaeps13_E1}
\end{figure}

The jumps of the permittivities  at the phase transition point $\Delta\varepsilon_{11}(E_{1,3})$ and $\Delta\varepsilon_{33}(E_{1,3})$ are nearly proportional to the squares of the strengths of the fields  $E_1$ and $E_3$ (figure~\ref{Deltaeps13_E1}):
\vspace{-4mm}
\bea &&
\Delta\varepsilon_{11}(E_{1})=k_{11}^{\varepsilon}E_{1}^{2},
\qquad
\Delta\varepsilon_{11}(E_{3})=k_{13}^{\varepsilon}E_{3}^{2}, \nonumber\\
&& \Delta\varepsilon_{33}(E_{1})=k_{31}^{\varepsilon}E_{1}^{2},
\qquad
\Delta\varepsilon_{33}(E_{3})=k_{33}^{\varepsilon}E_{3}^{2},
\nonumber \eea
where the coefficients are $k_{11}=0.064$~Km$^2$/MV$^2$,
$k_{13}=1.0$~Km$^2$/MV$^2$,
$k_{31}=0.725$~Km$^2$/MV$^2$,
$k_{33}=12.5$~Km$^2$/MV$^2$.

\begin{figure}[!b]
\begin{center}
\includegraphics[scale=0.8]{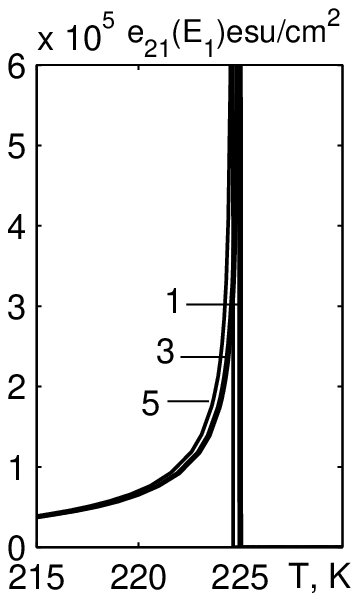}~~\includegraphics[scale=0.8]{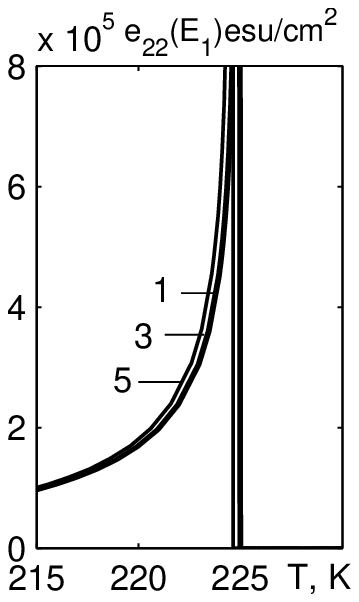}~~\includegraphics[scale=0.8]{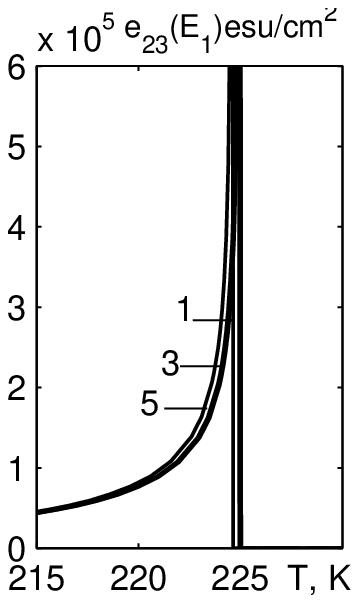}~~\includegraphics[scale=0.8]{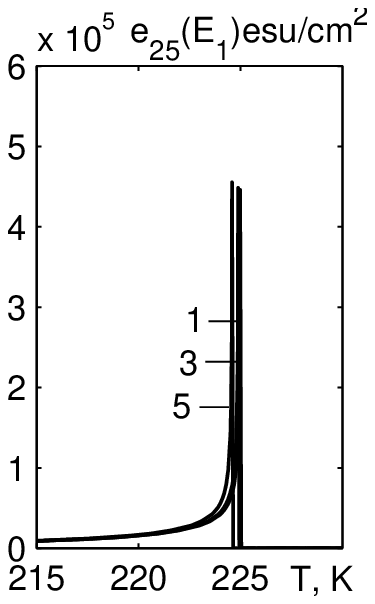}
\end{center}
 \vspace{-2mm}
\caption[]{ The temperature dependences of the coefficients of piezoelectric stress $e_{2i}$ of GPI crystal  at different values of the electric field $E_{1}$~(MV/m): 0.0~---~1;  2.0~---~3; 4.0~---~5. } \label{e21_E1}
\end{figure}

\begin{figure}[!t]
\begin{center}
\includegraphics[scale=0.8]{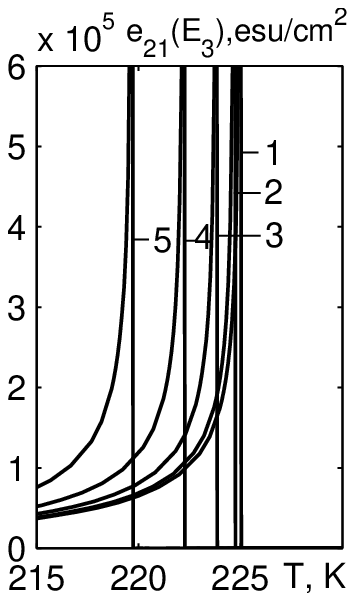}~~\includegraphics[scale=0.8]{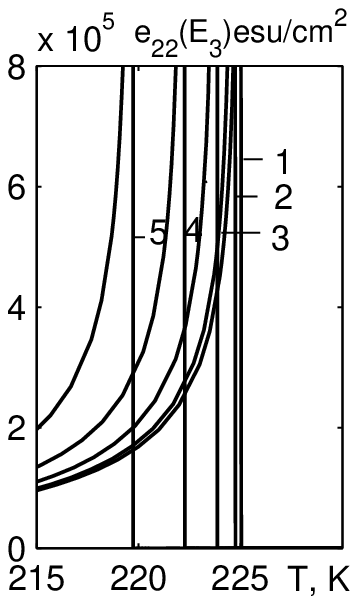}~~\includegraphics[scale=0.8]{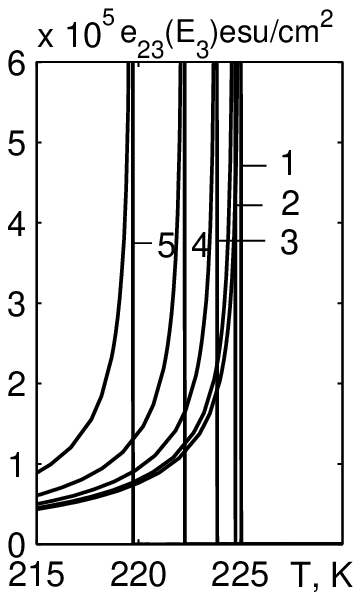}~~\includegraphics[scale=0.8]{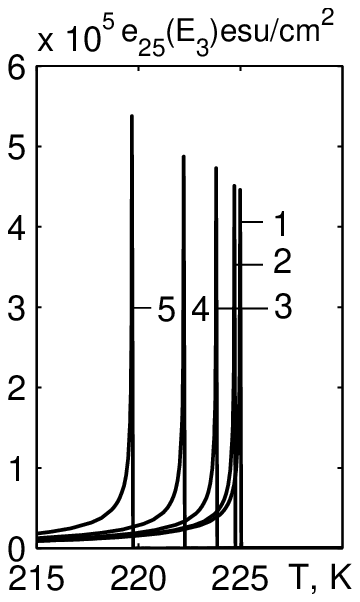}
\end{center}
\caption[]{ The temperature dependences of the coefficients of piezoelectric stress $e_{2i}$ of GPI crystal  at different values of the electric field $E_{3}$~(MV/m): 0.0~---~1; 1.0~---~2; 2.0~---~3; 3.0~---~4; 4.0~---~5. } \label{e21_E3}
\end{figure}

The temperature dependences of the coefficients of piezoelectric stress $e_{2i}$ at different values of the electric fields
 $E_1$ and $E_3$ are presented in figures~\ref{e21_E1} and~\ref{e21_E3}. An increase of the field  $E_1$ leads to a slight increase of piezomoduli $e_{2i}$. The splitting of the temperature dependences of $e_{2i}$ is much stronger in the case of field $E_3$.

\begin{figure}[!t]
\begin{center}
\includegraphics[scale=0.71]{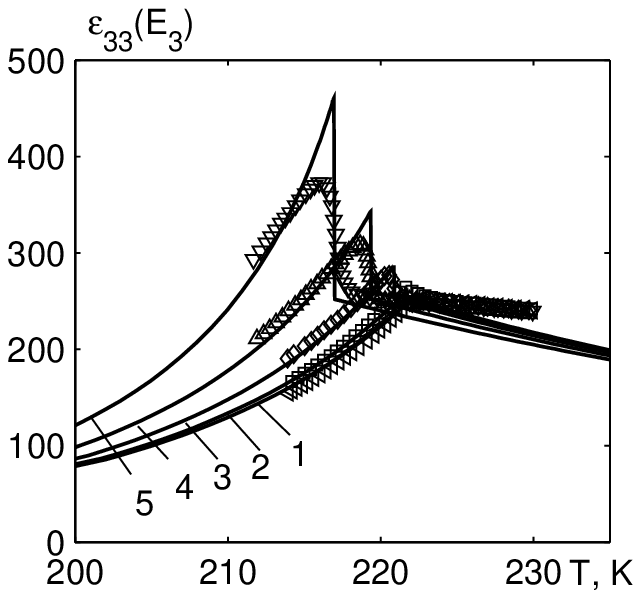}~~\includegraphics[scale=0.71]{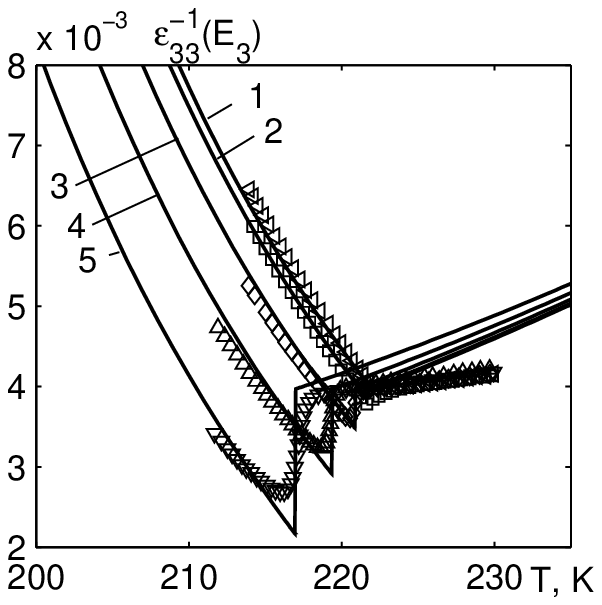}
\end{center}
\caption[]{ The temperature dependences of the static direct $\varepsilon_{33}$ and inverse $\varepsilon_{33}^{-1}$ permittivities of GPI crystal  at different values of the field  $E_3$~(MV/m): 0.0~---~1; 1.0~---~2; 2.0~---~3; 3.0~---~4; 4.0~---~5; symbols $\triangledown$, $\vartriangle$, $\lozenge$, $\square$, $\triangleleft$ are the experimental data  \cite{Stasyuk2003,Stasyuk2004}. } \label{eps33_E3_222}
\end{figure}

The results of an experimental  investigation of the static dielectric permittivity  $\varepsilon_{33}$ of GPI crystal at different values of the field $E_{3}$ are presented in \cite{Stasyuk2003,Stasyuk2004}. The phase transition temperature for this case was 222~K, but the field dependence of the  $T_{\text c}$ is similar to the crystal with  $T_{\text c}=225$~K.
Therefore, having made the above mentioned changes of the model parameters, we consider it possible to explain  the experimental data.
The calculated temperature dependences of the static direct $\varepsilon_{33}$ and inverse $\varepsilon_{33}^{-1}$ permittivities of GPI crystal  at different values of the field $E_{3}$ as well as the experimental data are presented in figure~\ref{eps33_E3_222}.
It is shown that at the phase transition temperature, theoretical curves $\varepsilon_{33}(T)$ have a sharp jump whose magnitude increases with an increase of the field.
However, the experimental curves $\varepsilon_{33}(T)$ are smooth, as in the case of a smeared phase transition.

\begin{figure}[!b]
\begin{center}
\includegraphics[scale=0.75]{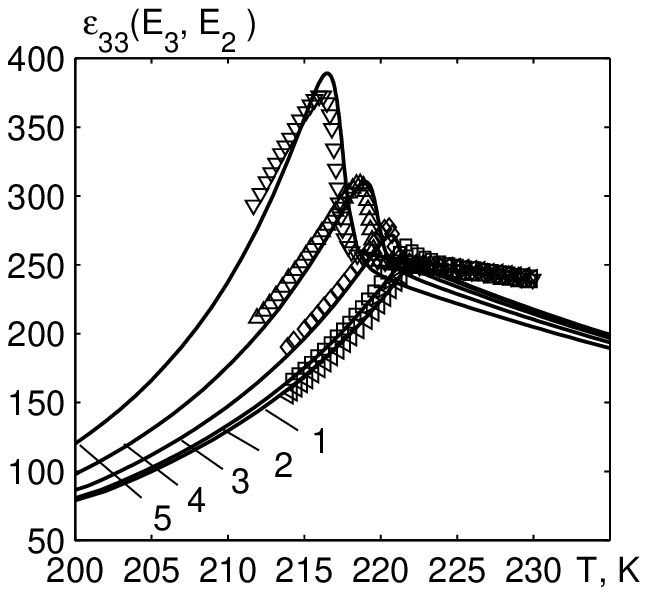}~~\includegraphics[scale=0.75]{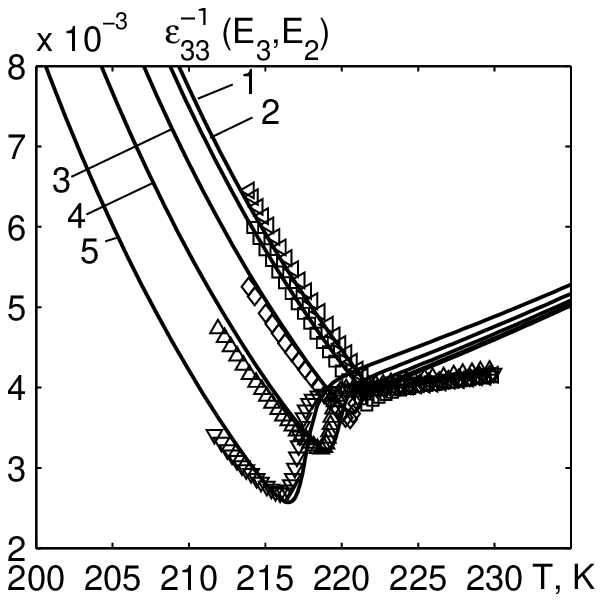}
\end{center}
\caption[]{The temperature dependences of the static direct $\varepsilon_{33}$ and inverse $\varepsilon_{33}^{-1}$ permittivities of GPI crystal  at different values of the field $E_3$~(MV/m): 0.0~---~1; 1.0~---~2; 2.0~---~3; 3.0~---~4; 4.0~---~5 and the field  $E_2=E_3/20$; symbols $\triangledown$, $\vartriangle$, $\lozenge$, $\square$, $\triangleleft$ are the experimental data \cite{Stasyuk2003,Stasyuk2004}. } \label{eps33_E3E2_222}
\end{figure}

In order to consider the reason of such a behaviour of permittivity $\varepsilon_{33}$, there was carried out a calculation of this component assuming that  together with the applied field $E_3$ there also appears an internal field $E_2$. As it turned out, one can achieve a satisfactory description of the temperature dependence of $\varepsilon_{33}$, assuming $E_2\sim0.05E_3$ (figure~\ref{eps33_E3E2_222}).
Such a component of the field $E_2$ could appear due to an incomplete reorientational relaxation of the glycine groups (which manifests itself during measurements in the hysteresis behaviour of $\varepsilon_{33}$); one cannot exclude the possibility of some deflection of the applied transverse field from the $OZ$-axis during the experiment (about $2.86\degree$). Nevertheless, if the effect is connected with the character and  peculiarities of internal fields in GPI crystal, the problem needs an additional study.

It also concerns the role of glycine groups in the phase transition in  GPI in the presence of external fields. Their deformation and reorientation is significant at the transition to the ferroelectric phase and manifests itself, for example,  in the experiment on Raman scattering \cite{Moreira2005} or at simulations of lattice dynamics \cite{Shchur2015}. At the same time, it was shown that the mechanism of the phase transition is connected with the proton ordering on hydrogen bonds.

It should be mentioned that the attempt to describe the behaviour of the inverse transverse dielectric permittivity at different electric fields within the phenomenological approach by means of Landau expansions was also done in \cite{Balashova2002,Balashova2007}. The authors explain the smeared minimum of the inverse permittivity below the transition temperature supposing that the  phase transition is of the first order one, close to the tricritical point. They qualitatively describe the experimental data  \cite{Stasyuk2003}, but quantitatively only at low fields. Such a supposition was based on their experimental data for GPI \cite{Balashova2002}, which noticeably differ from the obtained ones in the majority of other measurements. This can be connected with the unlike properties of the crystals grown at different conditions \cite{Balashova2007}.

\section{Conclusions}

Based on the proposed model of a deformed crystal, the calculation of dielectric characteristics of the crystal GPI in the presence of electric fields $E_1$ and $E_3$ is carried out. The obtained temperature and field dependences show that the effect of field $E_{3}$ on these characteristics is much more important than the effect of field $E_{1}$. At an increase of the field, the transition temperatures  $T_{\text c}(E_{1})$ and $T_{\text c}(E_{3})$ decrease almost as square of the field strengths. The magnitude  of the jumps of permittivities  $\varepsilon_{11}$ and $\varepsilon_{33}$ increases at the phase transition temperature according to the same law. Electric fields $E_{1}$ and $E_{3}$
cause polarizations  $P_{1}$ and~$P_{3}$; their temperature dependences are analyzed in the work.

The shape of anomalies of piezoelectric moduli in the region of a phase transition in the presence of transverse fields is analyzed. The obtained theoretical dependences have a character of predictions and can urge the subsequent experimental investigations.

At the same time, it is necessary to note that the ability  of GPI crystal to reorientate the local dipole moments and to change the orientation of the polarization vector by means of phase transition under reachable values of electric fields is unique.  We do not know any analogues among the ferroelectric crystals with hydrogen bonds.

Due to specific properties of GPI, special attention during investigations is also paid  to possible applications of the crystal in thin film structures  \cite{Balashova2013}; the role of impurities that introduce internal fields causing the appearance of pyroelectricity is studied \cite{Balashova2007}.

In our opinion, an important role is played by glycine ions that relatively easily change their orientations, exhibiting some inertia. Taking into account their relaxational dynamics, one could significantly supplement the comprehension of the mechanisms of external fields effect on dielectric properties of~GPI.

\section*{Acknowledgement}

The authors are indebted to Prof. Z. Czapla for helpful discussions and useful comments.

\appendix
\section{Parameters determining the local pseudospin susceptibilities with respect to electric fields and strains}
The notations introduced in equations~(\ref{X11})--(\ref{X33}) are as follows:

\renewcommand{\arraystretch}{1.2}
\begin{align}
\Delta&= \left| \begin{array}{cccc}
                2D -\varkappa_{11} & -\varkappa_{12} & -\varkappa_{13} & -\varkappa_{14} \\
                -\varkappa_{21} & 2D -\varkappa_{22} & -\varkappa_{23} & -\varkappa_{24} \\
                -\varkappa_{31} & -\varkappa_{32} & 2D -\varkappa_{33} & -\varkappa_{34}\\
                -\varkappa_{41} & -\varkappa_{42} & -\varkappa_{43} & 2D -\varkappa_{44}
                \end{array}
                \right|,\nonumber\\
 \Delta_{1}^{\chi \alpha}&= \left| \begin{array}{cccc}
                \varkappa_{1}^{\chi \alpha} & -\varkappa_{12} & -\varkappa_{13} & -\varkappa_{14} \\
                \varkappa_{2}^{\chi \alpha} & 2D -\varkappa_{22} & -\varkappa_{23} & -\varkappa_{24} \\
               \varkappa_{3}^{\chi \alpha} & -\varkappa_{32} & 2D -\varkappa_{33} & -\varkappa_{34}\\
                \varkappa_{4}^{\chi \alpha} & -\varkappa_{42} & -\varkappa_{43} & 2D -\varkappa_{44}
                \end{array}
                \right|,~~\Delta_{3}^{\chi \alpha}= \left| \begin{array}{cccc}
                2D -\varkappa_{11} & -\varkappa_{12} & \varkappa_{1}^{\chi \alpha} & -\varkappa_{14} \\
                -\varkappa_{21} & 2D -\varkappa_{22} & \varkappa_{2}^{\chi \alpha} & -\varkappa_{24} \\
                -\varkappa_{31} & -\varkappa_{32} & \varkappa_{3}^{\chi \alpha} & -\varkappa_{4}\\
                -\varkappa_{41} & -\varkappa_{42} & \varkappa_{4}^{\chi \alpha}& 2D -\varkappa_{44}
                \end{array}
                \right|,\nonumber\\
\Delta_{2}^{\chi \alpha}&= \left| \begin{array}{cccc}
                2D -\varkappa_{11} & \varkappa_{1}^{\chi \alpha} & -\varkappa_{13} & -\varkappa_{14} \\
                -\varkappa_{21} & \varkappa_{2}^{\chi \alpha} & -\varkappa_{23} & -\varkappa_{24} \\
                -\varkappa_{31} & \varkappa_{3}^{\chi \alpha} & 2D -\varkappa_{33} & -\varkappa_{34}\\
                -\varkappa_{41} & \varkappa_{4}^{\chi \alpha} & -\varkappa_{43} & 2D -\varkappa_{44}
                \end{array}
                \right|,~~\Delta_{4}^{\chi \alpha}= \left| \begin{array}{cccc}
                2D -\varkappa_{11} & -\varkappa_{12} & -\varkappa_{13} & \varkappa_{1}^{\chi \alpha} \\
                -\varkappa_{21} & 2D -\varkappa_{22} & -\varkappa_{23} & \varkappa_{2}^{\chi \alpha} \\
                -\varkappa_{31} & -\varkappa_{32} & 2D -\varkappa_{33} & \varkappa_{2}^{\chi \alpha}\\
                -\varkappa_{41} & -\varkappa_{42} & -\varkappa_{43} & \varkappa_{4}^{\chi \alpha}
                \end{array}
                \right|\label{A1},
   \end{align}
where
\begin{align}
&\varkappa_{f1}= \varkappa_{f;11}\varphi_{1}^{+}+\varkappa_{f;12}\beta\nu_{2}^{+}+\varkappa_{f;13}\varphi_{1}^{-}+\varkappa_{f;14}\beta\nu_{2}^{-}, \quad(f=1,2,3,4); \nonumber\\
&\varkappa_{f3}= \varkappa_{f;11}\varphi_{3}^{+}+\varkappa_{f;12}\beta\nu_{2}^{+}-\varkappa_{f;13}\varphi_{3}^{-}-\varkappa_{f;14}\beta\nu_{2}^{-}, \nonumber\\
&\varkappa_{f2}= \varkappa_{f;12}\varphi_{2}^{+}+\varkappa_{f;11}\beta\nu_{2}^{+}+\varkappa_{f;14}\varphi_{2}^{-}+\varkappa_{f;13}\beta\nu_{2}^{-}, \nonumber\\
&\varkappa_{f4}= \varkappa_{f;12}\varphi_{4}^{+}+\varkappa_{f;11}\beta\nu_{2}^{+}-\varkappa_{f;14}\varphi_{4}^{-}-\varkappa_{f;13}\beta\nu_{2}^{-}, \nonumber\\
&\varkappa_{f}^{\chi x}=\varkappa_{f;13}\beta\mu_{13}^{x}+\varkappa_{f;15}\beta\mu_{24}^{x}\,, \quad \varkappa_{f}^{\chi y}= \varkappa_{f;11}\beta\mu_{13}^{y}+\varkappa_{f;12}\beta\mu_{24}^{y}\,, \quad \varkappa_{f}^{\chi z}= \varkappa_{f;13}\beta\mu_{13}^{z}+\varkappa_{f;14}\beta\mu_{24}^{z}\,,\nonumber\\
&\varphi_{1,3}^{\pm}=\frac{1}{1 -  \eta_{1,3}^{2}} +\beta\nu_{1}^{\pm}=\frac{1}{1 -  \eta_{1,3}^{2}} +\frac{\beta}{4}(J_{11}\pm J_{13}),\nonumber\\
&\varphi_{2,4}^{\pm}=\frac{1}{1 -  \eta_{2,4}^{2}} +\beta\nu_{3}^{\pm}=\frac{1}{1 -  \eta_{2,4}^{2}} +\frac{\beta}{4}(J_{22}\pm J_{24}),\quad\beta\nu_{2}^{\pm}=\frac{\beta}{4}(J_{12}\pm J_{14}),\nonumber\\
&\nu_{1}^{\pm}=\nu_{1}^{0\pm}+\Bigg(\sum\limits_{i=1}^3\psi_{1i}^{\pm}\varepsilon_{i}\pm
\sum\limits_{j=4}^6\psi_{1j}^{\pm}\varepsilon_{j}\Bigg),\quad\nu_{1}^{0\pm}=\frac{1}{4}(J_{11}^{0}\pm J_{13}^{0});\quad\psi_{1i}^{\pm}=\frac{1}{4}(\psi_{11i}\pm\psi_{13i}),\nonumber\\
&\nu_{2}^{\pm}=\nu_{2}^{0\pm}+\Bigg(\sum\limits_{i=1}^3\psi_{2i}^{\pm}\varepsilon_{i}\pm
\sum\limits_{j=4}^6\psi_{2j}^{+}\varepsilon_{j}\Bigg),\quad\nu_{2}^{0\pm}=\frac{1}{4}(J_{12}^{0}\pm J_{14}^{0});\quad\psi_{2i}^{\pm}=\frac{1}{4}(\psi_{12i}\pm\psi_{14i}),\nonumber\\
&\nu_{3}^{\pm}=\nu_{3}^{0\pm}+\Bigg(\sum\limits_{i=1}^3\psi_{3i}^{\pm}\varepsilon_{i}\pm
\sum\limits_{j=4}^6\psi_{3j}^{\pm}\varepsilon_{j}\Bigg),\quad\nu_{3}^{0\pm}=\frac{1}{4}(J_{22}^{0}\pm J_{24}^{0});\quad\psi_{3i}^{\pm}=\frac{1}{4}(\psi_{22i}\pm\psi_{24i}),   \nonumber\\
&\varkappa_{1,3;11}= (l_{1+3}^{c}+l_{5+6}^{c})-\eta_{1,3}(l_{1+3}^{s}+l_{5+6}^{s}),\quad\varkappa_{1,3;12}= (l_{1-3}^{c}\mp l_{7-8}^{c})-\eta_{1,3}(l_{1-3}^{s}+l_{7+8}^{s}),\nonumber\\
&\varkappa_{1,3;13}=\pm(l_{2+4}^{c}+l_{7+8}^{c})-\eta_{1,3}(l_{2+4}^{s}-l_{7-8}^{s}),\quad\varkappa_{1,3;14}=(\pm l_{2-4}^{c}-l_{5-6}^{c})-\eta_{1,3}(l_{2-4}^{s}-l_{5-6}^{s}),\nonumber\\
&\varkappa_{2,4;11}= (l_{1-3}^{c}\mp l_{5-6}^{c})-\eta_{2,4}(l_{1+3}^{s}+l_{5+6}^{s}),\quad\varkappa_{2,4;12}= (l_{1+3}^{c}+l_{7+8}^{c})-\eta_{2,4}(l_{1-3}^{s}+l_{7+8}^{s}),\nonumber\\
&\varkappa_{2,4;13}= (\pm l_{2-4}^{c}- l_{7-8}^{c})-\eta_{2,4}(l_{2+4}^{s}-l_{7-8}^{s}),\quad\varkappa_{2,4;14}= (\pm l_{2+4}^{c}\pm l_{5+6}^{c})-\eta_{2,4}(l_{2-4}^{s}-l_{5-6}^{s}),\nonumber\\
&\varkappa_{1,3;15}=(\mp l_{2-4}^{c}+l_{5-6}^{c})-\eta_{1,3}(-l_{2-4}^{s}+l_{5-6}^{s}),\quad\varkappa_{2,4;15}=\mp( l_{2+4}^{c}+l_{5+6}^{c})+\eta_{2,4}(-l_{2-4}^{s}+l_{5-6}^{s}),\nonumber\\
&l_{1\pm3}^{c}= \cosh n_{1}\pm a^{2}\cosh n_{3};\quad l_{2\pm4}^{c}= \cosh n_{2}\pm a^{2}\cosh n_{4};\nonumber\\
&l_{5\pm6}^{c}= a\cosh n_{5}\pm a\cosh n_{6};\quad l_{7\pm8}^{c}= a\cosh n_{7}\pm a\cosh n_{8};\nonumber\\
&l_{1\pm3}^{s}= \sinh n_{1}\pm a^{2}\sinh n_{3};\quad l_{2\pm4}^{s}= \sinh n_{2}\pm a^{2}\sinh n_{4};\nonumber\\
&l_{5\pm6}^{s}= a\sinh n_{5}\pm a\sinh n_{6};\quad l_{7\pm8}^{s}= a\sinh n_{7}\pm a\sinh n_{8}.\label{A2}
\end{align}

The notations introduced in equations (\ref{e2l}) are as follows:
\begin{align}
\Delta_{1l}^{e}&= \left| \begin{array}{cccc}
                \varkappa_{1l}^{e} & -\varkappa_{12} & -\varkappa_{13} & -\varkappa_{14} \\
                \varkappa_{2l}^{e} & 2D -\varkappa_{22} & -\varkappa_{23} & -\varkappa_{24} \\
               \varkappa_{3l}^{e} & -\varkappa_{32} & 2D -\varkappa_{33} & -\varkappa_{34}\\
                \varkappa_{4l}^{e} & -\varkappa_{42} & -\varkappa_{43} & 2D -\varkappa_{44}
                \end{array}
                \right|,~~\Delta_{3l}^{e}= \left| \begin{array}{cccc}
                2D -\varkappa_{11} & -\varkappa_{12} & \varkappa_{1l}^{e} & -\varkappa_{14} \\
                -\varkappa_{21} & 2D -\varkappa_{22} & \varkappa_{2l}^{e} & -\varkappa_{24} \\
                -\varkappa_{31} & -\varkappa_{32} & \varkappa_{3l}^{e} & -\varkappa_{4}\\
                -\varkappa_{41} & -\varkappa_{42} & \varkappa_{4l}^{e}& 2D -\varkappa_{44}
                \end{array}
                \right|,\nonumber\\
\Delta_{2l}^{e}&= \left| \begin{array}{cccc}
                2D -\varkappa_{11} & \varkappa_{1l}^{e} & -\varkappa_{13} & -\varkappa_{14} \\
                -\varkappa_{21} & \varkappa_{2l}^{e} & -\varkappa_{23} & -\varkappa_{24} \\
                -\varkappa_{31} & \varkappa_{3l}^{e} & 2D -\varkappa_{33} & -\varkappa_{34}\\
                -\varkappa_{41} & \varkappa_{4l}^{e} & -\varkappa_{43} & 2D -\varkappa_{44}
                \end{array}
                \right|,~~\Delta_{4l}^{e}= \left| \begin{array}{cccc}
                2D -\varkappa_{11} & -\varkappa_{12} & -\varkappa_{13} & \varkappa_{1l}^{e} \\
                -\varkappa_{21} & 2D -\varkappa_{22} & -\varkappa_{23} & \varkappa_{2l}^{e} \\
                -\varkappa_{31} & -\varkappa_{32} & 2D -\varkappa_{33} & \varkappa_{2l}^{e}\\
                -\varkappa_{41} & -\varkappa_{42} & -\varkappa_{43} & \varkappa_{4l}^{e}
                \end{array}
                \right|\nonumber,\\
&\varkappa_{fl}^{e}=\beta(\psi_{1l}^{+}\varkappa_{f;11}+\psi_{2l}^{+}\varkappa_{f;12})(\eta_{1}+\eta_{3}) +
\beta(\psi_{2l}^{+}\varkappa_{f;11}+\psi_{3l}^{+}\varkappa_{f;12})(\eta_{2}+\eta_{4})\nonumber\\&\quad\quad+\beta(\psi_{1l}^{-}\varkappa_{f;13}+\psi_{2l}^{-}\varkappa_{f;14})
(\eta_{1}-\eta_{3})
+\beta(\psi_{2l}^{-}\varkappa_{f;13}+\psi_{3l}^{-}\varkappa_{f;14})(\eta_{2}-\eta_{4})+2\beta\delta_{l}(\rho_{f;1}+\rho_{f;2}),\nonumber\\
&\psi_{1l}^{\pm}=\frac{1}{4}(\psi_{11l}\pm\psi_{13l}),\quad\psi_{2l}^{\pm}=\frac{1}{4}(\psi_{12l}\pm\psi_{14l}),
\quad\psi_{3l}^{\pm}=\frac{1}{4}(\psi_{22l}\pm\psi_{24l}),
\nonumber\\
&\rho_{1,3;1}=-2(l^{s}_{3\pm4}-\eta_{1,3}l^{c}_{3+4}),\quad \rho_{1,3;2}=-l^{s}_{5+6}\pm l^{s}_{7-8}+\eta_{1,3}(l^{c}_{5+6}+l^{c}_{7+8}),\nonumber\\
&\rho_{2,4;1}=2(l^{s}_{3\pm4}+\eta_{2,4}l^{c}_{3+4}), \quad\rho_{2,4;2}=\pm l^{s}_{5-6}- l^{s}_{7+8}+\eta_{2,4}(l^{c}_{5+6}+l^{c}_{7+8}),\nonumber\\
&l_{3\pm4}^{s}=  a^{2}\sinh n_{3}\pm a^{2}\sinh n_{4}. \label{A3}
\end{align}

\newpage
\ukrainianpart

\title{Вплив електричних полів  на діелектричні    властивості   сегнетоелектрика   GPI}
\author{ І.Р. Зачек\refaddr{label1}, Р.Р. Левицький\refaddr{label2}, А.С. Вдович\refaddr{label2}, І.В. Стасюк\refaddr{label2}}
\addresses{\addr{label1} Національний університет ``Львівська політехніка'', вул. С. Бандери, 12, 79013 Львів, Україна
\addr{label2} Інститут фізики конденсованих систем НАН України, вул. Свєнціцького, 1, 79011 Львів, Україна
 }

\makeukrtitle

\begin{abstract}
\tolerance=3000%
Використовуючи модель GPI, модифіковану шляхом врахування п'єзоелектричного зв'язку з
деформаціями  $\varepsilon_i$ в наближенні двочастинкового кластера, розраховано компоненти вектора
поляризації та тензора статичної діелектричної проникності  кристала при прикладанні зовнішніх поперечних електричних полів $E_1$ і $E_3$.
Проведено аналіз впливу цих полів на діелектричні характеристики GPI.
При належному виборі параметрів теорії отримано  задовільний кількісний опис наявних експериментальних даних для цих характеристик.
\keywords сегнетоелектрики,  електричне поле, поляризація, діелектрична проникність, фазовий перехід
\end{abstract}

\end{document}